\documentclass[12pt]{article}%
\usepackage[nosort]{cite}
\usepackage{graphicx}
\usepackage{multicol}
\usepackage{amsfonts}
\usepackage{amssymb}
\usepackage{amsmath}
\usepackage{dsfont}
\usepackage{heck}
\usepackage{afterpage}
\usepackage{setspace}
\usepackage{verbatim}
\usepackage{color}
\usepackage{longtable}
\usepackage{float}
\usepackage{subcaption}
\usepackage{epsfig}
\usepackage{epstopdf}
\usepackage{adjustbox}
\usepackage{tikz}
\usepackage[margin=1in]{geometry}
\usepackage{titletoc}
\usepackage{hyperref}%
\providecommand{\U}[1]{\protect\rule{.1in}{.1in}}
\pdfoutput=1
\newsavebox{\mysavebox}

\hypersetup{colorlinks,citecolor=black,filecolor=black,linkcolor=black,urlcolor=black,pdftex}
\usetikzlibrary{decorations.markings}

\numberwithin{equation}{section}

\newcommand{\ba}{\begin{eqnarray}}
\newcommand{\ea}{\end{eqnarray}}

\newcommand{\be}{\begin{equation}}
\newcommand{\ee}{\end{equation}}

\tikzstyle{startstop} = [rectangle, rounded corners, minimum width=3cm, minimum height=1cm,text centered, draw=black, fill=blue!10]
\tikzstyle{startstop} = [rectangle, rounded corners, minimum width=3cm, minimum height=1cm,text centered, draw=black, fill=blue!10]
\tikzstyle{io} = [trapezium, trapezium left angle=70, trapezium right angle=110, minimum width=3cm, minimum height=1cm, text centered, draw=black, fill=blue!30]
\tikzstyle{process} = [rectangle, minimum width=3cm, minimum height=1cm, text centered, draw=black, fill=orange!30]
\tikzstyle{decision} = [diamond, minimum width=3cm, minimum height=1cm, text centered, draw=black, fill=green!30]
\tikzstyle{arrow} = [thick,->,>=stealth]
\begin{document}

\date{August 2016}

\title{Kinetic Mixing at Strong Coupling}

\institution{HARVARD}{\centerline{${}^{1}$ Jefferson Physical Laboratory, Harvard University, Cambridge, MA 02138, USA}}

\institution{UNC}{\centerline{${}^{2}$Department of Physics, University of North Carolina, Chapel Hill, NC 27599, USA}}

\institution{WOLFRAM}{\centerline{${}^{3}$Wolfram Research, Somerville, MA 02144, USA}}

\institution{GRUMPS}{\centerline{${}^{4}$Game Grumps and Ninja Sex Party, Los Angeles, CA 91436, USA }}

\authors{Michele Del Zotto\worksat{\HARVARD}\footnote{e-mail: {\tt delzotto@physics.harvard.edu}},
Jonathan J. Heckman\worksat{\UNC}\footnote{e-mail: {\tt jheckman@email.unc.edu}},\\[4mm]
Piyush Kumar\worksat{\WOLFRAM}\footnote{e-mail: {\tt piyush.kumar1@gmail.com}},
Arada Malekian\worksat{\UNC}\footnote{e-mail: {\tt aradam@email.unc.edu}},
and Brian Wecht\worksat{\GRUMPS}}

\abstract{A common feature of many string-motivated particle physics models is additional
strongly coupled $U(1)$'s. In such sectors, electric and magnetic states have comparable mass,
and integrating out modes also charged under $U(1)$ hypercharge generically
yields CP preserving electric kinetic mixing and CP violating magnetic kinetic mixing terms.
Even though these extra sectors are strongly coupled, we show that
in the limit where the extra sector has approximate $\mathcal{N} = 2$ supersymmetry,
we can use formal methods from Seiberg-Witten theory to compute these couplings.
We also calculate various quantities of phenomenological
interest such as the cross section for scattering between visible sector states and
heavy extra sector states, as well as the effects of supersymmetry breaking induced from coupling to the MSSM.}

\maketitle

\tableofcontents

\enlargethispage{\baselineskip}

\setcounter{tocdepth}{2}

\newpage

\section{Introduction \label{sec:INTRO}}

As the only known viable theory of quantum gravity, it is clearly important to
determine possible low energy manifestations of string theory. One promising
route to forging such connections is to examine generic string-motivated
scenarios for physics beyond the Standard Model of particle physics.

A generic feature of many string constructions is the presence of additional
$U(1)$ gauge fields. These can arise from dimensional reduction of higher
p-form potentials, that is, from the \textquotedblleft closed string
sector\textquotedblright\ of a model. Another common way such gauge fields
arise is from degrees of freedom localized on lower-dimensional branes, that
is, from the \textquotedblleft open string sector.\textquotedblright\ In many
cases, there can be degrees of freedom charged under both the $U(1)$
hypercharge factor of the Standard Model gauge group and one of these extra
$U(1)$'s. This motivates the study of kinetic mixing in the context of string
phenomenology. For a partial list of references, see e.g.,
\cite{Holdom:1985ag, Holdom:1986eq, Babu:1996vt, Dienes:1996zr, Lust:2003ky,
Abel:2003ue, Abel:2004rp, Abel:2006qt, Abel:2008ai, Benakli:2009mk,
Brummer:2009cs, Bruemmer:2009ky, Brummer:2009zg, Heckman:2010fh,
Heckman:2011sw, Goodsell:2011wn, Jaeckel:2012yz}, as well as
\cite{Cicoli:2011yh, Andreas:2011in, Andreas:2013iba, Jaeckel:2013ija,
Vogel:2013raa, Jaeckel:2014qea, Redondo:2015iea, Schwarz:2015lqa,
Kunze:2015noa, Vinyoles:2015khy, Arias:2016vxn}.

But another generic feature of many string constructions is the presence of
sectors which are strongly coupled \cite{Dine:1985he}. Indeed, while it is
certainly possible to arrange for \textit{some }parameters to remain weakly
coupled (as necessary for realizing the perturbative couplings of our world),
it is typically more problematic to arrange for \textit{all} couplings to be
small. In the context of closed string parameters, this is the statement that
it is easier --albeit less calculable-- to produce models with some geometric
moduli set at string scale values. In the case of open string sectors, this is
the statement that there are extra sectors at strong coupling.

Having such strongly coupled extra sectors is also expected to generate novel
phenomenological scenarios. For a review of some recent work on composite dark
matter with strong coupling dynamics, see for example \cite{Kribs:2016cew}.
Unparticles with a mass gap \cite{Strassler:2006im, Georgi:2007ek} provide
another class of strongly coupled extra sectors with novel signatures.

In this paper we combine these considerations, that is, we study
string-motivated scenarios with an extra $U(1)$ which is strongly coupled.
From this perspective, the gauge group of the Standard Model can be
approximated as a weakly gauged flavor symmetry. It is natural to expect there
to be states (which may be quite heavy) that are charged under both the
Standard Model and such extra $U(1)$'s. As far as we are aware, there have
been only limited analyses of such systems, with very specialized structure
for magnetic objects \cite{Brummer:2009cs, Bruemmer:2009ky}.

Kinetic mixing between a visible sector $U(1)$ and an extra sector $U(1)$ is
captured by the effective Lagrangian:
\begin{align}
L_{U(1)}  &  =L_{\text{diag}}+L_{\text{mix}}\\
L_{\text{diag}}  &  =-\frac{1}{4}F_{\mu\nu}F^{\mu\nu}-\frac{1}{4}F_{\mu\nu
}^{\prime}F^{\prime\mu\nu}+\frac{g^{2}\theta}{32\pi^{2}}F_{\mu\nu}^{\prime
}\widetilde{F}^{\prime\mu\nu}\\
L_{\text{mix}}  &  =-\frac{\chi_{\text{elec}}}{2}F_{\mu\nu}F^{\prime\mu\nu
}-\frac{\chi_{\text{mag}}}{2}F_{\mu\nu}\widetilde{F}^{\prime\mu\nu},
\end{align}
where $F_{\mu\nu}^{\prime}$ is the field strength of an extra $U(1)$ with
magnetic dual field strength $\widetilde{F}_{\mu\nu}^{\prime}=\frac{1}%
{2}\varepsilon_{\mu\nu\rho\sigma}F^{\prime\rho\sigma}$. Here, we have omitted
the theta angle of the visible sector since all its magnetic objects are
assumed to be quite heavy. The analogue of the fine structure constant in the
extra sector is $\alpha_{\mathrm{{extra}}} = g^{2} / 4 \pi$, so that strong
coupling corresponds to taking $\alpha_{\mathrm{{extra}}} \sim O(1)$.

A priori, then, kinetic mixing can occur via both a CP\ preserving and a CP
violating term:%
\begin{align}
\text{Electric Mixing}  &  \text{: }F_{\mu\nu}F^{\prime\mu\nu}\\
\text{Magnetic Mixing}  &  \text{: }F_{\mu\nu}\widetilde{F}^{\prime\mu\nu}.
\end{align}
Electric kinetic mixing has been heavily studied, starting with
\cite{Holdom:1985ag, Holdom:1986eq}, and has led to a slew of novel dark
matter scenarios. For some examples, see references \cite{Goldberg:1986nk,
Pospelov:2007mp, ArkaniHamed:2008qn, ArkaniHamed:2008qp, Baumgart:2009tn,
Cheung:2009qd, Cheung:2009su}.

\begin{figure}[ptb]
\centering
\includegraphics[
scale = 0.75, trim = 15mm 75mm 0mm 75mm
]{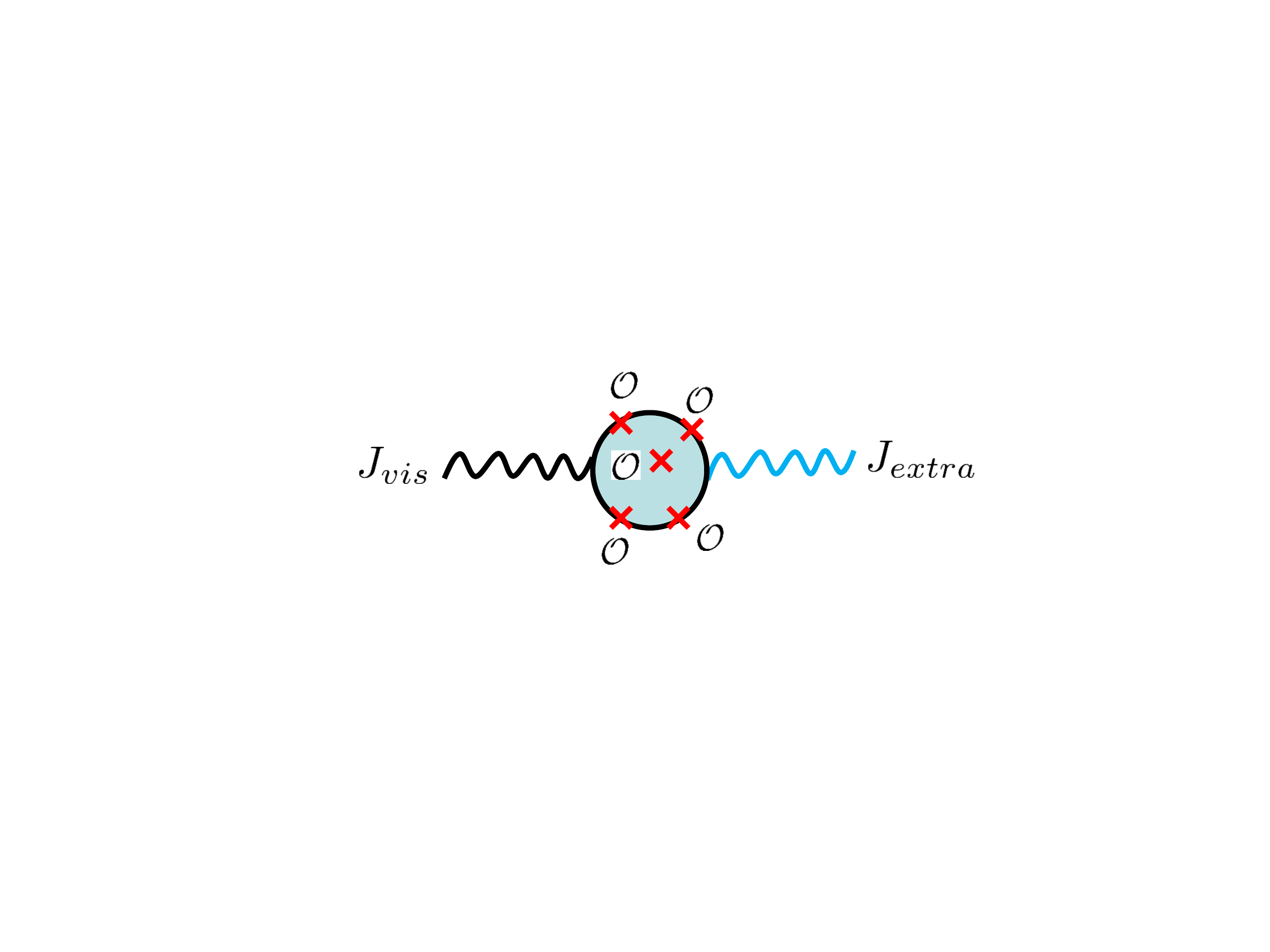}\caption{Depiction of kinetic mixing with a strongly coupled
extra sector. In this limit, the standard one loop calculation of kinetic
mixing does not apply and we must instead resort to non-perturbative methods.
Integrating out messenger states between the two sectors leads to electric and
magnetic kinetic mixing with the visible sector $U(1)$.}%
\label{MixoExample}%
\end{figure}Magnetic kinetic mixing is far more challenging to study. If we
have both electrically and magnetically charged states of comparable mass, we
are inherently at strong coupling, and there is no duality transformation
available to eliminate terms such as $F_{\mu\nu}\widetilde{F}^{\prime\mu\nu}$.
Indeed, another symptom of this fact is that when magnetic monopoles are
present, $F_{\mu\nu}\widetilde{F}^{\prime\mu\nu}$ can no longer be expressed
as a total derivative because there is no Lorentz invariant formulation of the
theory with a vector potential.\footnote{See, however, \cite{Schwarz:1993vs}.}
Indeed, it has been known for some time that the analogue of the QCD theta
angle plays an important role in the dynamics of abelian gauge theories with
dyons (i.e., states with electric and magnetic charge) \cite{Witten:1979ey}.

Precisely because the extra $U(1)$ is at strong coupling, standard methods
from perturbative quantum field theory do not apply. It is therefore important
to see whether we can extract \textit{any} quantitative information about
kinetic mixing at strong coupling.

In this paper we develop a general set of methods to extract these mixing
effects. In the limit where the extra sector enjoys approximate $\mathcal{N}%
=2$ supersymmetry, we show how to adapt formal methods from Seiberg-Witten
theory \cite{Seiberg:1994rs, Seiberg:1994aj} to extract the exact form of
electric and magnetic mixing. We also use these methods to extract the
spectrum of stable objects, and to calculate the leading order effects of
supersymmetry breaking induced from coupling to the MSSM. Additionally, we
calculate the leading order contributions to scattering between visible sector
states and heavy extra sector states. For some previous uses of extended
supersymmetry in the model building literature, see for example
\cite{Fox:2002bu} and for some discussion on other uses of magnetically
charged states of an extra sector, see e.g. \cite{Csaki:2010cs}.

In F-theory realizations of the Standard Model (see e.g.,
\cite{Heckman:2008rb, Heckman:2010bq, Weigand:2010wm} for reviews), the
canonical example of such an extra sector is a D3-brane probing a stack of
seven-branes with $E_{8}$ gauge symmetry \cite{Heckman:2010fh, Heckman:2011sw,
Heckman:2010qv , Heckman:2011hu, Heckman:2011bb, Heckman:2012nt,
Heckman:2012jm, Heckman:2015kqk}. That is, this realizes an $\mathcal{N}=2$
superconformal field theory with $E_{8}$ flavor symmetry \cite{Minahan:1996fg,
Minahan:1996cj}. Tilting the seven-branes and activating background fluxes
then breaks this flavor symmetry down to the Standard Model gauge group, which
in particular contains a $U(1)\subset E_{8}$ which we identify with
hypercharge of the Standard Model.

Approximate conformal symmetry of the extra sector means that the overall mass
scales of the extra sector are dictated by coupling it to additional sectors.
This can include both mass scales associated with the visible sector Standard
Model and its embedding in the MSSM\ and a stringy GUT, but can also include
other decoupled sectors (for example, in gravity mediated supersymmetry
breaking scenarios). For this reason, motivated values for approximate
$\mathcal{N}=2$ supersymmetric extra sector states can range from the TeV
scale up to the GUT scale. As noted in reference \cite{Heckman:2011sw},
partial breaking to $\mathcal{N}=1$ supersymmetry via T-brane deformations
\cite{Heckman:2010qv} can induce a seesaw like mechanism for dark exra sector
states, which in turn can generate sub-TeV mass scales.

We also put some of these considerations together to provide a preliminary
analysis of how such extra sectors can serve as toy models for more realistic
phenomenology. In particular, we explain how such extra sectors arise in
specific string constructions, and how to incorporate the leading order
effects of supersymmetry breaking. Since the resulting cosmological history
greatly depends on the associated mass scales, we mainly illustrate the general
contours of how such models work.

The rest of this paper is organized as follows. First, in section
\ref{sec:MagMix} we discuss in greater detail some additional features of
electric and magnetic mixing, as well as the effect such terms make on
scattering cross sections. Next, in section \ref{sec:SUSY} we show how to
apply formal methods from the study of theories with $\mathcal{N}=2$
supersymmetry to calculate such mixing effects, and how to incorporate the
leading effects of supersymmetry breaking. Section \ref{sec:RANKONE} sets up
the ingredients needed for theories with a single extra $U(1)$, which we
follow with an analysis of kinetic mixing when the extra sector is the rank
one $H_{1}$ Argyres-Douglas theory. In section \ref{sec:TOY} we discuss some
aspects of the resulting phenomenology. We present our conclusions in section
\ref{sec:CONC}. Some of the results presented in this paper also appear in the
PhD thesis of A. Malekian \cite{THESIS}.

\section{Electric and Magnetic Mixing \label{sec:MagMix}}

Our plan in this section will be to discuss some basic aspects of electric and
magnetic kinetic mixing. We also show how to extract some information about
how visible states can scatter off of dark dyons of an extra sector. For a
complementary account of some aspects of magnetic mixing, see for example
\cite{Brummer:2009cs, Bruemmer:2009ky}.

Our starting point is a system with $r$ total $U(1)$'s, with effective
Lagrangian:%
\begin{align}
L_{U(1)^{\prime}s} &  =\underset{i}{\sum}\left(  -\frac{1}{4}F^{i}\cdot
F^{i}+\frac{g_{ii}^{2}\theta_{ii}}{32\pi^{2}}F^{i}\cdot\widetilde{F}%
^{i}\right)  \\
&  +\underset{i\neq j}{\sum}\left(  -\frac{\chi_{ij}^{\text{elec}}}{4}%
F^{i}\cdot F^{j}-\frac{\chi_{ij}^{\text{mag}}}{4}F^{i}\cdot\widetilde{F}%
^{j}\right)  .
\end{align}
Working in terms of the electric and magnetic field strengths, we see two
types of interaction terms: Those which preserve CP\ and those which do not:%
\begin{align}
\text{CP Preserving} &  \text{: }F^{i}\cdot F^{j}\\
\text{CP Violating} &  \text{: }F^{i}\cdot\widetilde{F}^{j}\text{,}%
\end{align}
which are respectively associated with electric kinetic mixing and magnetic
kinetic mixing.

Now, since our extra $U(1)$'s will typically be at strong coupling, it is
actually more convenient to make use of a basis of fields in which charge
quantization is manifest. By abuse of notation, we shall use the same
expression for the field strengths:%
\begin{align}
L_{U(1)^{\prime}s} &  =-\frac{1}{4g_{ij}^{2}}F^{i}\cdot F^{j}+\frac
{\theta_{ij}}{32\pi^{2}}F^{i}\cdot\widetilde{F}^{j}\\
&  =-\frac{1}{16\pi}\left(  \operatorname{Im}\tau_{ij}F^{i}\cdot
F^{j} - \operatorname{Re}\tau_{ij}F^{i}\cdot\widetilde{F}^{j}\right)  ,
\end{align}
where we sum repeated indices, and we have introduced the complexified
parameter:%
\begin{equation}
\tau_{ij}=\frac{4\pi i}{g_{ij}^{2}}+\frac{\theta_{ij}}{2\pi}.
\end{equation}
The original mixing parameters are then given by:%
\begin{equation}
\chi_{ij}^{\text{elec}}=\frac{\operatorname{Im}\tau_{ij}}{\sqrt
{\operatorname{Im}\tau_{ii}}\sqrt{\operatorname{Im}\tau_{jj}}}\text{ \ \ and
\ \ }\chi_{ij}^{\text{mag}}=-\frac{\operatorname{Re}\tau_{ij}}{\sqrt
{\operatorname{Im}\tau_{ii}}\sqrt{\operatorname{Im}\tau_{jj}}}.\label{elecmag}%
\end{equation}

We are interested in extra sectors which contain both monopoles and dyons.
Some care must be taken in properly defining a basis of electric and magnetic
charges which is also consistent with Dirac quantization. It is convenient to
adopt a basis in which all magnetic charges are integral and in which the
physically measured electric charges may contain shifts by the various theta
angles \cite{Witten:1979ey}. So, we introduce $2r$ integers $n_{i}%
^{\text{elec}}$ and $n_{\text{mag}}^{i}$, and corresponding electric and
magnetic charges:%
\begin{equation}
Q_{i}^{\text{elec}}=\left(  n_{i}^{\text{elec}}-\frac{\theta_{ij}}{2\pi
}n_{\text{mag}}^{j}\right)  \text{ \ \ and \ \ }Q_{\text{mag}}^{i}%
=n_{\text{mag}}^{i}.\label{integrality}%
\end{equation}
In our conventions, the electric fields $\vec{E}^{i}$ and magnetic fields
$\vec{B}^{i}$ for a point particle with these integral values satisfy:
\begin{equation}
\vec{\nabla}\cdot\vec{E}^{i}=4\pi\delta^{3}(\vec{x})\times\left(  \frac
{1}{\operatorname{Im}\tau}\right)  ^{ij}n_{j}^{\text{elec}}\text{ \ \ and
\ \ }\vec{\nabla}\cdot\vec{B}^{i}=4\pi\delta^{3}(\vec{x})\times n_{\text{mag}%
}^{i}.
\end{equation}

Electric-Magnetic duality in this setting amounts to the collection of
transformations which preserve the form of the Dirac pairing. We can, without
loss of generality, adopt a basis in which the pairing $\Omega$ has the
block-diagonal form:%
\begin{equation}
\Omega=\left[
\begin{array}
[c]{cc}
& \mathbf{1}_{r\times r}\\
-\mathbf{1}_{r\times r} &
\end{array}
\right]  . \label{DiracPairing}%
\end{equation}
We shall sometimes write $\Omega_{IJ}$ with indices $I,J=1,...,2r$, i.e., the
index runs over both the electric and magnetic charges.

Non-trivial duality transformations are then captured by $2r\times2r$ matrices
$M$ with integer values subject to the condition:%
\begin{equation}
M^{T}\Omega M=\Omega,
\end{equation}
that is, the dualities are captured by $Sp(2r,\mathbb{Z})$ transformations. It
acts on the complexified parameter matrix $\tau_{ij}$ as:%
\begin{equation}
\tau\mapsto(A\tau+B)(C\tau+D)^{-1},
\end{equation}
where we have decomposed $M$ according to the block structure:%
\begin{equation}
M_{2r\times2r}=\left[
\begin{array}
[c]{cc}%
A_{r\times r} & B_{r\times r}\\
C_{r\times r} & D_{r\times r}%
\end{array}
\right]  \in Sp(2r,\mathbb{Z}).
\end{equation}

An important aspect of such duality transformations is that we must ensure
that our answers are compatible with this $Sp(2r,%
%TCIMACRO{\U{2124} }%
%BeginExpansion
\mathbb{Z}
%EndExpansion
)$ redundancy.\footnote{More precisely, it may happen that duality
transformations may only involve a congruent subgroup of $Sp(2r,%
%TCIMACRO{\U{2124} }%
%BeginExpansion
\mathbb{Z}
%EndExpansion
)$. This is in turn dictated by the precise spectrum of BPS\ objects which
transform into one another under various duality transformations. We shall not
dwell on this point in what follows. \label{notsp2r}} It is common to work in a
``fundamental domain'' for $\tau$, and label all charges with respect to this basis choice.
For the purposes of mapping out possible values of parameters, however, it is
sometimes convenient to work on the enlarged covering space. Unitarity imposes
the condition that:%
\begin{equation}
\operatorname{Im}\tau>0\text{,}%
\end{equation}
that is, that we have a positive definite matrix of kinetic terms. As we have
already remarked, this choice of parameterization contains some redundancies,
because we can also quotient by the duality group.

\subsection{Dark Rutherford Scattering}

Let us now suppose we have fixed a choice of fundamental domain, as well as a
basis of electric and magnetic charges. We would like to know how visible
sector states interact with hidden sector dyons.

The main idea will be to introduce a fixed background for our various fields.
We then consider small fluctuations around this background, which we identify
with the visible sector gauge potential. For this approximation to be valid,
we really need the extra sector states to be heavy, i.e., that we can simply
substitute in the background values of the various fields. This can be viewed
as a mild generalization of the calculation given in \cite{Coleman:1982cx}
(see also \cite{Bruemmer:2009ky}).

With this in mind, we shall aim to expand the various field strengths around
background values, with fluctuations captured by a vector potential:%
\begin{equation}
F_{\mu\nu}^{i}=F_{\mu\nu}^{i,\text{bkgnd}}+\partial_{\mu}A_{\nu}^{i}%
-\partial_{\nu}A_{\mu}^{i}. \label{EXPANDO}%
\end{equation}
Our goal will be to determine how the vector potentials $A_{\mu}^{i}$ couple
to the background sourced by a dyon. To proceed further, it is helpful to work
directly with the electric and magnetic field strengths. The mixing Lagrangian
is then given by:%
\begin{equation}
L_{U(1)^{\prime}s}=\frac{1}{2g_{ij}^{2}}\left(  \vec{E}^{i}\cdot\vec{E}%
^{j}-\vec{B}^{i}\cdot\vec{B}^{j}\right)  -\frac{\theta_{ij}}{8\pi^{2}}\vec
{E}^{i}\cdot\vec{B}^{j}.
\end{equation}
Since we are working with static pointlike sources, it suffices to consider
the coupling of the scalar potential to this background:%
\begin{equation}
\vec{E}^{i}=\vec{E}_{\text{bkgnd}}^{i}-\vec{\nabla}\varphi^{i}.
\end{equation}
Plugging in to our effective Lagrangian, the scalar potential couples to a
source term:%
\begin{equation}
J_{i}^{\text{eff}}=\delta^{3}(\vec{x})\times\left(  n_{i}^{\text{elec}%
}-\operatorname{Re}\tau_{ij}n_{\text{mag}}^{j}\right)  .
\end{equation}
Consequently, we see that in matrix elements between visible sector currents
and a heavy dark dyon, all our amplitudes will be proportional to the
quantity:%
\begin{equation}
\Pi(M_{\text{vis}},N_{\text{hid}})=q_{\text{vis}}\left(  \frac{1}%
{\operatorname{Im}\tau}\right)  ^{\text{vis,}j}\left(  n_{j}^{\text{elec}%
}-\operatorname{Re}\tau_{jk}n_{\text{mag}}^{k}\right)  ,
\end{equation}
in the obvious notation.

It is tempting to organize this into a single duality invariant expression.
Indeed, the scattering amplitude we compute cannot depend on the particular
basis of fields we choose to use in performing our calculation. The caveat is
that if we perform a duality transformation on the gauge fields and couplings,
we must also transform the charges of the external states entering into the
scattering amplitude.

So, following the discussion in \cite{Bruemmer:2009ky}, we note that the
$Sp(2r,%
%TCIMACRO{\U{2124} }%
%BeginExpansion
\mathbb{Z}
%EndExpansion
)$ invariant bilinear between dyonic charges is:%
\begin{equation}
\Pi(M,N)=M_{I}\Pi^{IJ}N_{J}\label{jmn}%
\end{equation}
where:%
\begin{equation}
M_{I}=\left[
\begin{array}
[c]{c}%
n_{i}^{\text{elec}}\\
n_{\text{mag}}^{i}%
\end{array}
\right]  ,\text{ \ \ }N_{J}=\left[
\begin{array}
[c]{c}%
n_{j}^{\text{elec}}\\
n_{\text{mag}}^{j}%
\end{array}
\right]  ,
\end{equation}
and:%
\begin{equation}
\Pi^{IJ}=\left[
\begin{array}
[c]{cc}%
\left(  \frac{1}{\operatorname{Im}\tau}\right)  ^{il} & -\left(  \frac
{1}{\operatorname{Im}\tau}\right)  ^{ij}\operatorname{Re}\tau_{jl}\\
-\operatorname{Re}\tau_{ij}\left(  \frac{1}{\operatorname{Im}\tau}\right)
^{jl} & \operatorname{Re}\tau_{ij}\left(  \frac{1}{\operatorname{Im}\tau
}\right)  ^{jk}\operatorname{Re}\tau_{kl}+\operatorname{Im}\tau_{il}%
\end{array}
\right]  ,
\end{equation}
in the obvious notation. We view $M_{I}\Pi^{IJ}N_{J}$ as calculating the
matrix element between a visible sector current associated with $M_{I}$ and a
hidden sector current associated with $N_{J}$.

Consider, then, the special case where we have a state with charge $M_{I}$
which couples to a weakly coupled gauge boson, i.e., this is our
\textquotedblleft visible sector.\textquotedblright\ Assuming the extra sector
state is quite heavy and that the visible sector state has mass $m_{\text{vis}%
}$ and charge $q_{\text{vis}}$ and moves with velocity $\vec{v}$, we then get
a mild generalization of the standard result for Rutherford scattering (see
e.g. \cite{Peskin:1995ev}):%
\begin{equation}
\frac{d\sigma}{d\Omega}=\frac{\vert\Pi(M,N) \vert^{2}}{4m_{\text{vis}}%
^{2}v^{4}\sin^{4}\frac{\theta}{2}}. \label{skitskat}%
\end{equation}
An interesting feature of this formula is the dependence of the cross section
on electric and magnetic charge of the extra sector. In particular, we see
that the strength of the magnetic mixing term can have a non-trivial impact on
scattering of dark magnetic states.

We caution that to really apply this formula, we need to have at least one
scattering state to be near the free field limit, i.e., we need it to be
charged with respect to only weakly coupled gauge boson, and for the states of
the extra sector to be heavy. Thankfully, this is the case of maximal interest
for phenomenology, where we consider a visible sector electron / charged
nucleon scattering off of a heavy hidden sector dyon.

It is also convenient to package the contribution to the scattering amplitude
in terms of an effective electric charge from the extra sector. We define an
effective electric charge for a dark sector state which scatters off a visible
sector state:%
\begin{equation}
q_{\text{eff}}\equiv\frac{\left\vert \Pi(M_{\text{vis}},N_{\text{hid}%
})\right\vert }{\left\vert M_{\text{vis}}\right\vert \alpha_{\text{vis}}}.
\label{qeffdef}%
\end{equation}
Note that since $\Pi$ is linear in $M_{\text{vis}}$, the overall value of the
visible sector charge drops out of this expression.

One might also ask whether we can extend this calculation to a regime in which
we do not treat the extra sector as a fixed classical source. This is of
particular relevance for strongly coupled sectors where we can typically
expect a rich spectrum of composite bound states. When we do this, we need to
have much more detailed information about the spectrum of asymptotic
scattering states. It is analogous to the problem in QCD of determining the
precise form of the parton distribution functions. Nevertheless, we can
already see that several novel features will present themselves in the general
case. Precisely because we expect a general theory of dyons to include
non-trivial bound states with a finite radius, these configurations can have
non-trivial angular momentum (as dictated by the Dirac pairing). This already
tells us that if we consider a scattering event in which the internal state of
the composite object undergoes a transition, conservation of angular momentum
will lead to non-trivial selection rules on possible interaction terms. One
can view this as a generalization of the Callan-Rubakov effect
\cite{Rubakov:1982fp, Callan:1982ah, Callan:1982au}.

\section{Supersymmetric Approximation \label{sec:SUSY}}

In the previous section we presented some general considerations on electric
and magnetic mixing, and explained how in the regime where the dark charged
objects are quite heavy, we can determine the net effect of magnetic mixing on
the visible sector. In particular, many of the same considerations used to
study electric kinetic mixing also carry over to this case as well.

This prompts the question: Can we realize specific examples in which magnetic
mixing is generated, and moreover, can we actually \textit{calculate} the
overall strength of such mixing terms? To frame the discussion to follow, let
us recall that in a weakly coupled theory, the leading order contribution to
kinetic mixing between two $U(1)$'s is:%
\begin{equation}
\frac{1}{g_{ij}^{2}}=\underset{\psi}{\sum}c^{(\psi)}\frac{q_{i}^{(\psi)}%
q_{j}^{(\psi)}}{16\pi^{2}}\log\left(  \frac{M_{(\psi)}^{2}}{\mu^{2}}\right)  ,
\end{equation}
where the sum is over states of mass $M_{(\psi)}$, and the $q$'s are the
electric charges under the respective gauge groups. Additionally, $c^{(\psi)}$
is a numerical pre-factor which depends on the spin of the state.

We would like to generalize this calculation to the case where our extra
sector states interact with a strongly coupled $U(1)$. The issue we face is
that perturbative methods via Feynman diagrams will no longer apply.

To give specific examples of how to integrate out massive dyonic states to
calculate possible mixing terms, we shall use the general formalism of
supersymmetric gauge theories. Our conventions follow \cite{Wess:1992cp}.
Recall that in a supersymmetric gauge theory, we can package the
$\mathcal{N}=1$ vector multiplet (with a gauge field and its superpartner the
gaugino as dynamical degrees of freedom) in terms of the superfield
$W_{\alpha}=-i\lambda_{\alpha}(y)+...$. In this context, the electric and
magnetic mixing terms both descend from a single complexified parameter:%
\begin{equation}
\tau_{ij}=\frac{4\pi i}{g_{ij}^{2}}+\frac{\theta_{ij}}{2\pi},
\end{equation}
and the kinetic term is:%
\begin{align}
L_{U(1)^{\prime}s}  &  =\underset{i,j}{\sum}\operatorname{Im}\int d^{2}%
\theta\text{ }\frac{\tau_{ij}}{8\pi}W^{(i)}\cdot W^{(j)}\\
&  =\underset{i,j}{\sum}-\frac{1}{4g_{ij}^{2}}F^{(i)}\cdot F^{(j)}%
+\frac{\theta_{ij}}{32\pi^{2}}F^{(i)}\cdot\widetilde{F}^{(j)}.
\end{align}

In spite of this canonical holomorphic structure, it is still challenging to
extract the parameters $\tau_{ij}$ for a theory with both electric and
magnetically charged states, even with $\mathcal{N}=1$ supersymmetry. To
proceed further, we now assume that we have $\mathcal{N}=2$ supersymmetry. Let
us hasten to add that this will not require us to extend the $U(1)_{\text{vis}%
}$ gauge theory to actually have $\mathcal{N}=2$ supersymmetry. All that is
really required is that all extra sector states organize into $\mathcal{N}=2$
supersymmetry multiplets. Indeed, we shall view the visible sector as a weakly
gauged flavor symmetry.

Let us review some basic aspects of $N=2$ supersymmetric gauge theory. For
further details, see for example \cite{AlvarezGaume:1996mv, Bilal:2001nv}.
Now, an $\mathcal{N}=2$ vector multiplet consists of an $\mathcal{N}=1$ vector
multiplet and an $\mathcal{N}=1$ chiral multiplet. In our conventions, the
scalar component of each $\mathcal{N}=2$ vector multiplet is $a^{i}$. When the
$a^{i}$ have generic values, all states charged under the $U(1)$'s will have
picked up a mass and we can integrate them out.\footnote{For example, for a
weakly coupled $U(1)$ gauge theory in which we have an $\mathcal{N}=2$
hypermultiplet with electric charge $q_{\text{elec}}$, we have a
superpotential coupling $W=\sqrt{2}H^{c}(q_{\text{elec}}a)H.$ So, giving a
background value to the scalar $a$ gives a mass to the corresponding
hypermultiplet.} In this limit, then, we get a low energy effective action
involving $\mathcal{N}=2$ abelian vector multiplets. The key point for us is
that the parameters $\tau_{ij}$ are given by:%
\begin{equation}
\frac{\partial a_{i}^{D}}{\partial a^{j}}=\tau_{ij},
\end{equation}
where we have introduced the scalar of the magnetic dual theory $a_{i}^{D}$
given by the derivative of the $\mathcal{N}=2$ pre-potential:\footnote{Recall
that in terms of $\mathcal{N}=2$ superfields (which by abuse of notation we
also denote by $a^{i}$ and $a_{i}^{D}$), the low energy effective Lagrangian
specifies the pre-potential $\mathcal{F}\left(  a^{i}\right)  $ via:%
\begin{equation}
L_{\text{eff}}=\frac{1}{8\pi}\operatorname{Im}\int d^{2}\theta d^{2}%
\widetilde{\theta}\left(  \mathcal{F}\left(  a^{i}\right)  -a^{i}a_{i}%
^{D}\right)  .
\end{equation}
}%
\begin{equation}
\frac{\partial\mathcal{F}}{\partial a^{i}}=a_{i}^{D}.
\end{equation}
An additional benefit of knowing the specific values of these parameters is
that we can also extract the mass $M$ of BPS states with prescribed electric
and magnetic charges from the central charge $Z$. For a state of charge
$Q^{I}=(q_{\text{elec}}^{1},...,a_{\text{elec}}^{r};q_{1}^{\text{mag}%
},...,q_{r}^{\text{mag}})$ which also transforms in a representation
$\mathcal{R}$ of a flavor symmetry $G_{\text{flav}}$, we have:%
\begin{equation}
Z=\underset{I,J}{\sum}\Omega_{IJ}Q^{I}A^{J}+\frac{1}{\sqrt{2}}%
\underset{b=1}{\overset{\text{dim}\mathcal{R}}{\sum}}q_{b}m^{b}\text{,\ with
\ \ }M^{2}=2\left\vert Z\right\vert .
\end{equation}
Here, we have introduced $A^{J}=(a_{1}^{D},...,a_{r}^{D};a^{1},...,a^{r})$,
which pair with the charges via the Dirac pairing $\Omega_{IJ}$ of equation
(\ref{DiracPairing}). We have also introduced background mass parameters
$m_{b}$ which transform in the representation $\mathcal{R}$ along with
corresponding half integrally quantized charges:%
\begin{equation}
q_{b}\in\frac{1}{2}%
%TCIMACRO{\U{2124} }%
%BeginExpansion
\mathbb{Z}
%EndExpansion
.
\end{equation}
Physically, we should view the mass parameters as being specified by weakly
gauging a flavor symmetry and moving onto the Coulomb branch. From this
perspective, we introduce a complex scalar $\phi$ in the adjoint
representation of $G_{\text{flav}}$. Activating a vev for this field yields a
mass for the hypermultiplet:\footnote{For example, in a weakly coupled model,
with a hypermultiplet in a representation $\mathcal{R}$, we have the
superpotential coupling $\sqrt{2}H^{c}T_{A}^{\mathcal{R}}\phi^{A}H$, where
$T_{A}^{\mathcal{R}}$ are generators of $G_{\text{flav}}$ in the
representation $\mathcal{R}$.}%
\begin{equation}
\frac{m^{b}}{\sqrt{2}}=\vec{w}^{b}\cdot\vec{\phi},\label{massparams}%
\end{equation}
where $\vec{w}^{b}$ is a weight vector for a representation $\mathcal{R}$ of
$G_{\text{flav}}$, and:%
\begin{equation}
\vec{\phi}=\underset{s=1}{\overset{\text{rk }G}{\sum}}\vec{\alpha}_{s}\phi
^{s},
\end{equation}
where the $\vec{\alpha}_{s}$'s are a basis of positive roots of the flavor symmetry algebra.

To extract the kinetic mixing with a visible sector, as well as the mass of
various electric and magnetic states, our task therefore reduces to computing
$a_{i}^{D}$ as a function of the values $a^{j}$ and the $\phi^{s}$. In
particular, if we identify one of the flavor $U(1)$'s with the visible sector
$U(1)$ so that $\phi^{\text{vis}}=a^{\text{vis}}$, we can extract the kinetic
mixing term:%
\begin{equation}
\tau_{\text{vis},i}=\frac{\partial a_{j}^{D}}{\partial\phi^{\text{vis}}}%
=\frac{\partial a_{j}^{D}}{\partial a^{\text{vis}}}.
\end{equation}

Thankfully, this is precisely what the general method outlined by Seiberg and
Witten in \cite{Seiberg:1994rs, Seiberg:1994aj} provides. The key point for us
is that there is an auxiliary Riemann surface and a meromorphic one-form
$\lambda$ (i.e., a one-form with simple poles) such that the parameters
$a^{i}$, $a_{i}^{D}$ and $m^{b}$ are encoded as contour integrals
\cite{Seiberg:1994rs, Seiberg:1994aj}. The presence of marked points can be
visualized as the effect of weakly gauging a $U(1)$, i.e. adding a long narrow
tube to the Seiberg-Witten curve.

\subsection{Supersymmetry Breaking Effects \label{ssec:BREAK}}

A priori, it could happen that even if supersymmetry is badly broken in the
visible sector, it may be preserved in some approximate form in the extra
sector. Indeed, the primary assumption we make throughout this work is the
presence of (possibly mildly broken) $\mathcal{N}=2$ supersymmetry in the
extra sector. The nature of supersymmetry breaking will of course impact some
details of the mass spectrum, as well as the amount of mixing between the
visible and hidden sectors. Our aim here will therefore be to focus on aspects
which are more generic. In particular, we focus on those contributions which
come from coupling to the visible sector.

Since we are working in the limit where we treat the visible sector as a
weakly gauged symmetry, we can parameterize possible contributions in terms of
non-zero background values to the corresponding $\mathcal{N}=2$ vector
multiplet. Assuming these effects are small, we can expand in their auxiliary
fields. In terms of $\mathcal{N}=1$ superfields $a^{\text{vis}}$ and
$W^{\text{vis}}$, we can therefore make the subsitutions:%
\begin{equation}
a^{\text{vis}}\mapsto a^{\text{vis}}+\theta^{2}F^{\text{vis}}\text{ \ \ and
\ \ }W_{\alpha}^{\text{vis}}\mapsto W_{\alpha}^{\text{vis}}+\theta_{\alpha
}D^{\text{vis}}.\label{expander}%
\end{equation}
For example, F-term breaking could arise from a symmetry breaking pattern
which also breaks a GUT\ group to the Standard Model gauge group. D-term
breaking will inevitably arise in the MSSM\ and its extensions due to the
D-term potential of the MSSM. Expanding as in line (\ref{expander}) is valid
provided these mass scales are sub-dominant compared with supersymmetric mass
terms:%
\begin{equation}
\frac{F^{\text{vis}}}{M_{\text{SUSY}}^{2}}\ll1\text{\ \ and \ \ }%
\frac{D^{\text{vis}}}{M_{\text{SUSY}}^{2}}\ll1.
\end{equation}

Let us begin by tracking the impact of F-term supersymmetry breaking on the
hypermultiplets. First of all, we can see that the BPS\ mass formula will now
receive corrections. To see why, note that the mass of a hypermultiplet with
electric-magnetic charge vector $Q^{I}$ and flavor charges $q_{b}$ has central
charge:%
\begin{equation}
Z_{Q,q}(a,a^{D},m)=\underset{I,J}{\sum}\Omega_{IJ}Q^{I}A^{J}+\frac{1}{\sqrt
{2}}\underset{b}{\sum}q_{b}m^{b}.
\end{equation}
In particular, a hypermultiplet $H^{c}\oplus H$ has a superpotential coupling:%
\begin{equation}
L_{eff}\supset\int d^{2}\theta\sqrt{2}H^{c}Z_{Q,q}H+h.c..
\end{equation}
Expanding around the background of line (\ref{expander}), we get:%
\begin{equation}
Z_{Q,q}\mapsto Z_{Q,q}+\theta^{2}F^{\text{vis}}\frac{\partial Z_{Q,q}%
}{\partial a^{\text{vis}}}.
\end{equation}
If this is the only effect of supersymmetry breaking, we can calculate the
correction to the masses of states in the hypermultiplets:%
\begin{equation}
\left\vert M_{\pm}^{\text{Bosons}}\right\vert ^{2}=2\left\vert Z_{Q,q}%
\right\vert ^{2}\pm\sqrt{2}\left\vert F^{\text{vis}}\frac{\partial Z_{Q,q}%
}{\partial a^{\text{vis}}}\right\vert \text{ \ \ and \ \ }\left\vert
M^{\text{Fermions}}\right\vert ^{2}=2\left\vert Z_{Q,q}\right\vert
^{2}.\label{splitter}%
\end{equation}
This approximation requires $F^{\text{vis}}/M_{\text{SUSY}}^{2}\ll1$. Observe
also that the lightest state in the hypermultiplet is a boson, and that the
supertrace relation on the mass spectrum is obeyed.

An interesting feature of this answer is that there are actually two distinct
contributions to the mass splitting formula. First, we have the expected
electric contribution from the mass parameters proportional to $q_{b}m^{b}$.
For a magnetically charged state, there is another contribution\ proportional
to $\partial a^{D}/\partial a^{\text{vis}}=\tau_{\text{mix}}$.

Consider next the effects of D-term supersymmetry breaking on the vector
multiplets. To track these contributions, we return to our kinetic mixing
interactions, and make the substitution of line (\ref{expander}):%
\begin{equation}
L_{eff}\supset\frac{1}{8\pi}\left(  \operatorname{Im}\left(  \int d^{2}%
\theta\tau_{ij}W^{(i)}\cdot W^{(j)}\right)  +\operatorname{Im}\left(
F^{\text{vis}}\frac{\partial\tau_{ij}}{\partial a^{\text{vis}}}\lambda
^{(i)}\cdot\lambda^{(j)}\right)  +\operatorname{Im}\left(  \tau_{\text{vis,}%
j}D^{(\text{vis})}\cdot D^{(j)}\right)  \right)  .
\end{equation}
The middle term induces a gaugino mass matrix, which in particular can mix a
visible sector gaugino with the extra sector gauginos. The last term specifies
an effective FI\ parameter for the extra sector \cite{Baumgart:2009tn,
Cheung:2009qd}.

The net combination of contributions, in particular the presence of
FI\ parameters and mass terms for the hypermultiplets provides multiple ways
in which supersymmetry may be partially or fully broken. First of all, in the
$\mathcal{N}=2$ supersymmetric limit, we see that in various weakly coupled
models, having a large mass but with an\ FI\ parameter switched on will lead
to a partial breaking of $\mathcal{N}=2$ to $\mathcal{N}=1$ supersymmetry
\cite{Antoniadis:1995vb}. Additionally, in this case the vacuum generically
sits at the origin of the Coulomb branch and one of the scalars of the
hypermultiplet develops a vev, breaking the $U(1)$, thus screening some
charges (the ones which are local with respect to the hypermultiplet
charge)\ and confining others (the ones which are non-local with respect to
the hypermultiplet charge). In such cases, we do not expect to retain as much
analytic control, because $\mathcal{N}=2$ supersymmetry has been badly broken.

An alternative way to retain more analytic control is to also introduce a
superpotential mass term for the Coulomb branch scalar of the extra sector. In
the context of string constructions where the extra sector originates from a
D3-brane probing a visible sector, this will generically happen in the
presence of appropriate fluxes / instanton effects \cite{Heckman:2010fh,
Heckman:2011sw, Heckman:2010qv, Cecotti:2009zf, Martucci:2006ij,
Marchesano:2009rz}. Provided the mass of the Coulomb branch scalar is lower
than that of the hypermultiplets, but still higher than the supersymmetry
breaking terms from mixing with the visible sector, we can continue to use the
$\mathcal{N}=2$ supersymmetric approximation developed here.

\section{Rank One Theories \label{sec:RANKONE}}

In this section we turn to some concrete examples of kinetic mixing at strong
coupling. For simplicity, we consider rank one theories, i.e., those with a
single $U(1)$ in the extra sector Coulomb branch.

We further specialize to extra sectors which are obtained from a deformation
of a strongly coupled $\mathcal{N}=2$ superconformal field theory with a
flavor symmetry group $G_{\text{flav}}$. This case is particularly
well-motivated from string constructions, as it arises from a probe D3-brane
next to a stack of intersecting seven-branes with exceptional gauge symmetry.
In such examples, the Standard Model is realized via the stack of
seven-branes, and the D3-brane realizes an extra sector \cite{Heckman:2010fh,
Heckman:2011sw, Heckman:2010qv, Heckman:2011hu,
Heckman:2011bb, Heckman:2012nt, Heckman:2012jm, Heckman:2015kqk}. We can describe these
theories as $\mathcal{N}=1$ deformations of $\mathcal{N}=2$ superconformal
field theories with exceptional flavor symmetry \cite{Minahan:1996fg,
Minahan:1996cj}. Our discussion of the associated $\mathcal{N}=2$
Seiberg-Witten geometry follows the presentation and analysis of reference
\cite{Noguchi:1999xq}.

We assume that $U(1)_{\text{vis}}$ corresponds to a weakly gauged subgroup of
$G_{\text{flav}}$. There can potentially be additional weakly gauged $U(1)$'s
contained in $G_{\text{flav}}$. We therefore denote the local electric and
magnetic coordinates as $a$ and $a^{D}$, and the various mass parameters as
$m^{b}$. The central charge of a state with electric charge $n_{\text{elec}}$
and magnetic charge $n_{\text{mag}}$ transforming in a representation
$\mathcal{R}$ of $G_{\text{flav}}$ is:%
\begin{equation}
Z=n_{\text{elec}}a-n_{\text{mag}}a^{D}+\frac{1}{\sqrt{2}}%
\underset{b=1}{\overset{\dim\mathcal{R}}{\sum}}q_{b}m^{b}\text{\ ,\ with
\ \ }M^{2}=2\left\vert Z\right\vert ^{2}.\label{Zentral}%
\end{equation}
The vacua are parameterized by $u$, the coordinate on the moduli space of
vacua. In physical terms, $u$ is given by the vacuum expectation value (vev)
of an operator of the strongly coupled field theory. A non-zero value for this
operator breaks conformal symmetry and gives masses to the hypermultiplets of
the theory.\footnote{In a weakly coupled $SU(2)$ gauge theory, it would be
given by Tr $\phi^{2}$, where $\phi$ is the adjoint valued scalar of the
$\mathcal{N}=2$ vector multiplet. In the case of a strongly coupled theory,
this characterization is not available. One symptom of this is that for the
$H_{1}$ Argyres-Douglas theory \cite{Argyres:1995jj}, for example, the scaling
dimension of $u$ is $4/3$, and for the $E_{8}$ Minahan-Nemeschansky theory
\cite{Minahan:1996fg, Minahan:1996cj} it has scaling dimension $6$. These
scaling dimensions are calculated using the method given in
\cite{Argyres:1995xn} (see also \cite{Aharony:2007dj}).} The corresponding
Seiberg-Witten curve is given by:
\begin{equation}
y^{2}=x^{3}+f(u,m)x+g(u,m).\label{cubic}%
\end{equation}
The coefficients $f$ and $g$ are determined by our choice of a strongly
coupled theory.

Let us now turn to the Seiberg-Witten differential. In general, we need to
introduce a meromorphic one-form with appropriate periods which captures the
spectrum of dyonic states in our theory. In fact, there can be more than one
choice, and this is dictated by picking a representation $\mathcal{R}$ for the
flavor symmetry group, so we denote the Seiberg-Witten differential by
$\lambda_{\mathcal{R}}$. Physically, however, the coupling constants will not
depend on this choice. In more formal terms, we are specifying a section of
the elliptic fibration over the $u$-plane. The general form of $\lambda
_{\mathcal{R}}$ is:%
\begin{equation}
\lambda_{\mathcal{R}}=\alpha\frac{xdx}{y}+\beta\frac{dx}{y}+\underset{b}{\sum
}\gamma_{b}y_{b}\frac{dx}{y(x-x_{b})},\label{SWdiff}%
\end{equation}
where the coefficients $\alpha$, $\beta$ and $\gamma_{b}$ depend on the
parameters $u$ and $m$. Here, $y_{b}$ is the value of $y$ in equation
(\ref{cubic}) evaluated at the point $x=x_{b}$. The parameters of the
effective action are in turn obtained by evaluating the contour integrals:%
\begin{equation}
a=\underset{\gamma_{A}}{%
%TCIMACRO{\doint }%
%BeginExpansion
{\displaystyle\oint}
%EndExpansion
}\lambda_{\mathcal{R}}\text{, \ \ }a^{D}=\underset{\gamma_{B}}{%
%TCIMACRO{\doint }%
%BeginExpansion
{\displaystyle\oint}
%EndExpansion
}\lambda_{\mathcal{R}}\text{, \ \ }\frac{1}{k_{\mathcal{R}}}\frac{m^{b}%
}{2\sqrt{2}}=\underset{x_{b}}{%
%TCIMACRO{\doint }%
%BeginExpansion
{\displaystyle\oint}
%EndExpansion
}\lambda_{\mathcal{R}},\label{contourintegrals}%
\end{equation}
where we have introduced mass parameters $m^{b}$ of the weakly gauged flavor
symmetry. These $m^{b}$ transform in the representation $\mathcal{R}$. Here,
$k_{\mathcal{R}}=$ Ind$(\mathcal{R})/n$ with Ind$(\mathcal{R})$ the index of
the representation, which in our conventions is set to $1$ for the fundamental
representation. Additionally, the parameter $n=1$ if all mass parameters are
associated with a unique pole $x_{b}$, and $n=2$ if each pole $x_{b}$ is
associated with two mass parameters. The additional factor of $1/2$ in the
last contour integral is due to the fact that we have a two sheeted Riemann
surface, but are only encircling the pole on one of the sheets.

Physically, the $x_{b}$ are marked points associated to long narrow cylinders
(i.e. weakly gauged flavor symmetries) and where $\gamma_{A}$ and $\gamma_{B}$
are a basis of one-cycles on the Riemann surface such that:%
\begin{equation}
\gamma_{A}\cap\gamma_{B}=1\text{, \ \ }\gamma_{A}\cap\gamma_{A}=0\text{,
\ \ }\gamma_{B}\cap\gamma_{B}=0\text{.}%
\end{equation}

Now, our aim is to calculate the kinetic mixing couplings of our model. To
this end, we will need to evaluate the derivatives:%
\begin{equation}
\tau_{\text{extra}}\equiv\frac{\partial a^{D}}{\partial a}\text{ \ \ and
\ \ }\tau_{\text{mix}}\equiv\frac{\partial a^{D}}{\partial a^{\text{vis}}},
\end{equation}
where $a^{\text{vis}}$ is the local coordinate of the visible sector Coulomb
branch, associated with the weakly gauge visible sector $U(1)$. In this
approximation we also have $\tau_{\text{vis}}\simeq i\infty$. Electric /
Magnetic duality in the strongly coupled $U(1)$ is the geometric statement
that there is in general an ambiguity in defining which one-cycle of our curve
is $\gamma_{A}$, and which is $\gamma_{B}$. A duality invariant way to
parameterize the strength of the extra sector coupling is in terms of the
Klein invariant $J$-function:%
\begin{equation}
J(\tau_{\text{extra}})=\frac{4f^{3}}{4f^{3}+27g^{2}},
\end{equation}
which satisfies $J(i)=1$ and $J(e^{2\pi i/6})=0$.

Our plan in the remainder of this section will be to illustrate how to
calculate the explicit form of these mixing terms. We first present the
expressions for the period integrals. We will need these in order to extract
numerical quantities of interest. After this, we turn to a concrete model
which exhibits strong coupling. We calculate the electric and magnetic kinetic
mixing parameters in this model, and also determine the spectrum of lightest
stable charged objects. One can view this as defining an interesting
phenomenological scenario in its own right, though from the perspective of a
complete string theory construction, it is better viewed as a toy model.

\subsection{Elliptic Integrals}

Since our eventual aim is to extract numerical values of the magnetic mixing,
we will need explicit expressions for the contour integrals of line
(\ref{contourintegrals}). Following \cite{Ferrari:1996de, AlvarezGaume:1997fg,
Bilal:1997st}, we introduce a basis of three elliptic integrals which we use
to express the contour integrals of the Seiberg-Witten differential around the
one-cycles of the Seiberg-Witten curve. In addition to the contours encircling
the poles, we have one-cycles which encircle the roots of the cubic in $x$
appearing in equation (\ref{cubic}):%
\begin{equation}
y^{2}=x^{3}+fx+g=(x-e_{1})(x-e_{2})(x-e_{3}),
\end{equation}
where the roots of the cubic are:%
\begin{equation}
e_{i}=-\frac{1}{\xi^{i-1}}\left(  \frac{2}{3\Lambda}\right)  ^{1/3}f+\frac
{\xi^{i-1}}{3}\left(  \frac{3\Lambda}{2}\right)  ^{1/3}\text{ \ \ with
\ \ }\Lambda=-9g+\sqrt{3}\sqrt{4f^{3}+27g^{2}}\text{ \ \ and \ \ }\xi=e^{2\pi
i/3}.
\end{equation}
We take a basis in which for $u$ and $m$ real, the cycle $\gamma_{A}$ is given
by encircling $e_{2}$ and $e_{3}$, and the cycle $\gamma_{B}$ encircles
$e_{1}$ and $e_{2}$. Using the presentation in \cite{Bilal:1997st}, we have
the explicit form of the contour integrals in terms of elliptic integrals:%
\begin{align}
I_{A}^{(1)} &  =\underset{\gamma_{A}}{%
%TCIMACRO{\doint }%
%BeginExpansion
{\displaystyle\oint}
%EndExpansion
}\frac{dx}{y}=\frac{4}{(e_{1}-e_{3})^{1/2}}K(k)\\
I_{A}^{(2)} &  =\underset{\gamma_{A}}{%
%TCIMACRO{\doint }%
%BeginExpansion
{\displaystyle\oint}
%EndExpansion
}\frac{xdx}{y}=\frac{4}{(e_{1}-e_{3})^{1/2}}\left[  e_{1}K(k)+(e_{3}%
-e_{1})E(k)\right]  \\
I_{A}^{(3)}(c) &  =\underset{\gamma_{A}}{%
%TCIMACRO{\doint }%
%BeginExpansion
{\displaystyle\oint}
%EndExpansion
}\frac{dx}{y(x-c)}=\frac{4}{(e_{1}-e_{3})^{3/2}}\left[  \frac{1}%
{1-\widetilde{c}+p}K(k)+\frac{4p}{1+p}\frac{1}{(1-\widetilde{c})^{2}-p^{2}}%
\Pi_{1}\left(  \nu(c),\frac{1-p}{1+p}\right)  \right]
\end{align}
with:%
\begin{equation}
k^{2}=\frac{e_{2}-e_{3}}{e_{1}-e_{3}}\text{, \ \ }p^{2}=\frac{e_{2}-e_{1}%
}{e_{3}-e_{1}}\text{, \ \ }\widetilde{c}=\frac{c-e_{3}}{e_{1}-e_{3}}\text{,
\ \ }\nu(c)=-\left(  \frac{1-\widetilde{c}+p}{1-\widetilde{c}-p}\right)
^{2}\left(  \frac{1-p}{1+p}\right)  ^{2}.
\end{equation}
Similar considerations hold for the integrals around $\gamma_{B}$ by
interchanging $e_{1}$ and $e_{3}$.

In the above, we have introduced the elliptic integrals (see e.g.
\cite{Erdelyi}):%
\begin{align}
K(k) &  =\underset{0}{\overset{1}{\int}}\frac{dx}{\left[  (1-x)^{2}%
(1-k^{2}x^{2})\right]  ^{1/2}}\\
E(k) &  =\underset{0}{\overset{1}{\int}}dx\left(  \frac{1-k^{2}x^{2}}{1-x^{2}%
}\right)  ^{1/2}\\
\Pi_{1}(\nu,k) &  =\underset{0}{\overset{1}{\int}}\frac{dx}{\left[
(1-x)^{2}(1-k^{2}x^{2})\right]  ^{1/2}(1+\nu x^{2})},
\end{align}
which in \texttt{Mathematica} are respectively given by $K(k)=$
\texttt{EllipticK[}$k^{2}$\texttt{]}, $E(k)=$ \texttt{EllipticE[}$k^{2}%
$\texttt{]}, \ $\Pi_{1}\left(  \nu,k\right)  =$ \texttt{EllipticPi[}%
$-\nu,k^{2}$\texttt{]}.

In obtaining numeric results, we must be mindful of a few subtleties. First of
all, the actual period integral expressions will depend on a basis of electric
and magnetic charges for the visible and extra sector. This can lead to shifts
in the evaluation of period integrals by contributions proportional to
$m/\sqrt{2}$. Our guiding principle is that we recover the correct asymptotics
for all periods and masses in suitable decoupling limits.

An additional subtlety has to do with the specific implementation in
\texttt{Mathematica}. In the numerical evaluation of these expressions we will
encounter branch cuts in the roots of the cubic in $x$. To account for this,
we fix one patch of values of the parameters for $m~$real and for small phases
of $u$, and then continue to other values by permuting the roots of the cubic
to retain smooth behavior for all numerically evaluated quantities.

\subsection{The $H_{1}$ Argyres-Douglas Theory \label{sec:AD}}

We now turn to a detailed analysis in the case where the extra sector is a
deformation of the $H_{1}$ Argyres-Douglas theory \cite{Argyres:1995jj}. This
is also sometimes referred to as the \textquotedblleft$A_{3}$ Argyres-Douglas
theory\textquotedblright\ because of the way it is engineered by taking type
IIB\ string theory on the background $\mathbb{R}^{3,1}\times X$, where $X$ is
a non-compact Calabi-Yau threefold with a local $A_{3}$ singularity
\cite{Eguchi:1996ds,Eguchi:1996vu,Shapere:1999xr}.

This is a four-dimensional $\mathcal{N}=2$ superconformal field theory which
enjoys an $SU(2)$ flavor symmetry.\footnote{The name $H_{1}$ simply comes from
the fact that in an F-theory construction of this model, we have a D3-brane
probing a non-perturbative bound state of $(p,q)$ seven-branes with $SU(2)$
flavor symmetry. Indeed, in F-theory there are two distinct ways to realize an
$SU(2)$ gauge symmetry on a seven-brane, one which is perturbative and is
called $A_{1}$ (realized by a type $I_{2}$ fiber), and one which is
non-perturbative, and is called $H_{1}$ (realized by a type $III$ fiber). For
additional discussion on this point, see e.g. \cite{Morrison:1996pp} and
\cite{Gaberdiel:1997ud}.} Now, in this theory, there is a single $U(1)$
subalgebra of $SU(2)$, so we have our Coulomb branch parameter $u$ and a
single complex scalar parameterizing breaking patterns of the flavor symmetry.
It therefore suffices to introduce mass parameters $m^{1}$ and $m^{2}$
transforming in the doublet representation. Returning to equation
(\ref{massparams}), we have:%
\begin{equation}
\frac{m^{1}}{\sqrt{2}}=\phi\text{ \ \ and \ \ }\frac{m^{2}}{\sqrt{2}}=-\phi,
\end{equation}
so we can work in terms of a single mass parameter $m=\sqrt{2}\phi$.

The Seiberg-Witten curve and Seiberg-Witten differential in the fundamental
representation are (see reference \cite{Noguchi:1999xq}):%
\begin{align}
y^{2}  &  =x^{3}+ux+w_{2}\\
\lambda &  =\frac{\sqrt{2}}{4\pi i}\left(  \frac{u}{3}+\frac{m_{1}y_{1}}%
{x}\right)  \frac{dx}{y}%
\end{align}
where:
\begin{equation}
w_{2}=-4m_{1}m_{2}=4m^{2} \label{wtwo}%
\end{equation}
is the mass dependent quadratic Casimir, and $y_1 = 2m_1$. Physically, one can view $w_2$
as the gauge invariant operator proportional to Tr$\phi^{2}$ we would get from weakly
gauging the $SU(2)$ flavor symmetry. In the above, we used equation
(\ref{contourintegrals}) with $k_{\mathcal{R}}=1/2$ (since we have the SW
differential in the fundamental representation, but there is a single pole).

As a first step towards understanding the parameter space of our model, we
compute the Klein-Invariant $J$-function:
\begin{equation}
J(\tau_{\text{extra}})=\frac{4u^{3}}{4u^{3}+27(4m^{2})^{2}}.
\end{equation}
So depending on the parameters, we can either be at strong coupling or weak
coupling. For example, three canonical values of interest are:%
\begin{align}
\tau_{\text{extra}} &  =i\text{ \ \ \ \ \ \ \ \ \ for \ \ }m=0\\
\tau_{\text{extra}} &  =e^{2\pi i/6}\text{\ \ \ \ for \ \ }u=0\\
\tau_{\text{extra}} &  \simeq i\infty\text{ \ \ \ \ \ \ for \ \ }\left(
\frac{u}{3}\right)  ^{3}+4m^{4}=0.\label{weakcoup}%
\end{align}
The parameters $u$ and $m$ each implicitly specify mass scales. More
precisely, because the $H_{1}$ Argyres-Douglas theory is (at the origin of
moduli space)\ actually a superconformal field theory, homogeneity allows us
to fix the scaling of $u$ and $m$ as a function of energy scales. We have:%
\begin{equation}
u\sim\text{Mass}^{4/3}\text{ \ \ and \ \ }m\sim\text{Mass.}\label{massscales}%
\end{equation}
The fractional power in the scaling of $u$ is one of the hallmarks of a
strongly coupled superconformal field theory. This leaves us with one unfixed
dimensionless ratio, $m^{4}/u^{3}$.

Depending on the phenomenological scenario, the actual mass scales involved
could be anywhere from the GUT\ scale down to the TeV\ or sub-TeV scale. For
example, in many string-motivated scenarios, it is natural to take
$m\sim10^{16}$ GeV since this is the implicit scale set by separating the
various seven-branes from each other. On the other hand, if we assume that the
dominant contribution to conformal symmetry breaking is set by supersymmetry
breaking effects, a far lower reference scale is also possible.

We now turn to the calculation of the periods $a$ and $a^{D}$ and their
derivatives. We have:%
\begin{align}
a  &  =\frac{\sqrt{2}}{4\pi i}\left(  \frac{2u}{3}I_{A}^{(1)}+w_{2}I_{A}%
^{(3)}(0)\right)  -\frac{2}{3}\frac{m}{\sqrt{2}}\\
a^{D}  &  =\frac{\sqrt{2}}{4\pi i}\left(  \frac{2u}{3}I_{B}^{(1)}+w_{2}%
I_{B}^{(3)}(0)\right)  +\frac{2}{3}\frac{m}{\sqrt{2}},
\end{align}
with $w_{2}=4m^{2}$, as per equation (\ref{wtwo}). Let us make a few comments
on the presence of the terms proportional to $m$ in our period integrals.
Strictly speaking, this last piece is just an artifact of how we pick a basis
of contour integrals, i.e., how we choose to define our basis of electric
charges with respect to the visible sector. The choice in the above equation
comes from imposing the condition that as we take the $m\rightarrow\infty$
decoupling limit, $a$ and $a^D$ should be independent of $m$. Additionally,
we pass to a theory with no continuous flavor symmetry, and
in which the asymptotic value of $a^{D}/a\rightarrow\exp(2\pi i/6)$, i.e., the
value of $\tau_{\text{extra}}$ in this limit is frozen. This induces a flow from the $H_1$
Argyres-Douglas theory to what is known as the $H_0$ Argyres-Douglas theory.

The first derivatives of the periods provide us with the complexified gauge
coupling and the mixing parameter:%
\begin{equation}
\tau_{\text{extra}}\equiv\frac{\partial a^{D}}{\partial a}=\frac{\partial
a^{D}/\partial u}{\partial a/\partial u}\text{ \ \ and \ \ }\tau_{\text{mix}%
}\equiv\frac{\partial a^{D}}{\partial\phi}=\sqrt{2}\frac{\partial a^{D}%
}{\partial m}.
\end{equation}
Taking one more derivative provides us with the terms which appear in gaugino
mixing (after supersymmetry breaking) as well as the coupling between the
Coulomb branch scalar and the $U(1)$ gauge fields. In evaluating these
derivatives, we must treat $a$ and $\phi$ as independent variables.

To get a sense of the overall values of these coupling constants, and to
emphasize the point that these really are calculable quantities, we present a
few of the numerically evaluated derivatives obtained via our method. In
general, it is challenging to obtain a set of parameters which remains in a
single fundamental domain (i.e., a single basis of electric and magnetic
charges). To bypass these issues and get a sensible class of
examples, we hold fixed fixed $u=0.1$, with $m$ in
powers of $5$. For the first derivatives of $a^{D}$, we have:%
\begin{equation}%
\begin{tabular}
[c]{|c|c|c|}\hline
& $\tau_{\text{extra}}=\partial a^{D}/\partial a$ & $\tau_{\text{mix}%
}=\partial a^{D}/\partial\phi$\\\hline
$m=0.04$ & $0.13+0.99i$ & $\left(  -2.0+1.0i\right)  \times10^{-1}$\\\hline
$m=0.2$ & $0.45+0.90i$ & $(-2.4+3.5i)\times10^{-2}$\\\hline
$m=1.0$ & $0.50+0.87i$ & $(-2.6+4.4i)\times10^{-3}$\\\hline
$m=5.0$ & $0.50+0.87i$ & $(-3.0+5.2i)\times10^{-4}$\\\hline
\end{tabular}
\ \ ,
\end{equation}
and for the second derivatives of $a^{D}$, we have:%
\begin{equation}%
\begin{tabular}
[c]{|c|c|c|c|}\hline
& $\partial^{2}a^{D}/\partial a\partial a$ & $\partial^{2}a^{D}/\partial
a\partial\phi$ & $\partial^{2}a^{D}/\partial\phi\partial\phi$\\\hline
$m=0.04$ & $3.6\times10^{0}-5.4i$ & $3.7-2.4i$ & $\left(  5.1-1.4i\right)
\times10^{0}$\\\hline
$m=0.2$ & $3.1\times10^{-1}-3.4i$ & $-0.7-2.1i$ & $\left(  2.4-2.9i\right)
\times10^{-1}$\\\hline
$m=1.0$ & $6.9\times10^{-3}-0.65i$ & $-0.2-0.4i$ & $\left(  4.9-8.0i\right)
\times10^{-3}$\\\hline
$m=5.0$ & $1.6\times10^{-4}-0.13i$ & $-0.05-0.08i$ & $\left(  1.1-1.9i\right)
\times10^{-4}$\\\hline
\end{tabular}
\ \ .
\end{equation}

Let us stress that the physically more meaningful quantity is given by a
duality invariant expression such as a scattering amplitude, as in our
discussion in section \ref{sec:MagMix}. The reason is that to get a proper
notion of the overall strength of kinetic mixing, we also need to know the
spectrum of charges in the extra sector which can couple to the visible sector.

\subsubsection{BPS\ Spectrum}

For various model building considerations it is important to know the spectrum
of stable objects in our system, and their charges in some duality frame under
both the extra sector $U(1)$, and the visible sector $U(1)$. In more realistic
models where supersymmetry is broken, the spectrum will be deformed with a
mass splitting specified as in our discussion around equation (\ref{splitter}%
). A non-zero mass splitting within a multiplet also means that there can now
be non-trivial decays to the lowest mass state. With an unbroken $U(1)$,
however, this bottom component will be stable. We therefore view the
$\mathcal{N}=2$ supersymmetric approximation as telling us the leading order
structure of stable objects in our theory.

Let us now turn to the BPS\ spectrum of the $H_{1}$ Argyres-Douglas theory.
With the explicit form of the period integrals in hand, we can also determine
the lightest BPS\ particles at any point on the Coulomb branch. For early work
on the BPS\ spectrum of Argyres-Douglas theories, see reference
\cite{Shapere:1999xr}. In general terms, the spectrum of stable BPS\ states in
the system will depend on the value of the Coulomb branch parameter and mass
parameters of the model. An additional feature is that we should expect
\textquotedblleft wall-crossing phenomena\textquotedblright\ in which the
spectrum of stable objects actually changes as we cross real codimension one
loci in the moduli space of vacua \cite{Seiberg:1994rs, Cecotti:1992qh,
Kontsevich:2008fj}.

Returning to the BPS formula for the mass of our states given in equation
(\ref{Zentral}), we have for a state of the rank one $H_{1}$ Argyres-Douglas
theory:%
\begin{equation}
Z=n_{\text{elec}}a-n_{\text{mag}}a^{D}+q_{\text{flav}}\frac{m}{\sqrt{2}%
}\text{\ ,\ with \ \ }M^{2}=2\left\vert Z\right\vert ^{2},
\end{equation}
so we see that if we take $m\rightarrow\infty$, a state with non-zero charge
with respect to the flavor symmetry will develop a large mass.

It is also possible to arrange for the flavor neutral state to be lightest by
appropriately tuning the parameters and moduli of the theory. For example, we
can ensure that we have an approximately massless state by working in the
special limit where the discriminant is nearly zero:%
\begin{equation}
4u^{3}+27(w_{2})^{2}\simeq0,
\end{equation}
with $w_{2}=4m^{2}$ given by equation (\ref{wtwo}). Indeed, in this case the
length of the cycle used to generate the period $a$ collapses to zero size,
and the corresponding BPS\ mass of a $U(1)_{\text{extra}}$ electrically
charged state will be zero. In the special case of the $H_{1}$ Argyres-Douglas
theory, we can also see that when $u$ and $m$ are both non-zero, the coupling
constant $\tau_{\text{extra}}$ will be near $i\infty$, i.e., the point of weak coupling.

Let us now turn to the calculation of the BPS\ spectrum of the theory in the
Coulomb phase. There are by now various methods for performing such a
calculation. These include the method of \textquotedblleft
BPS\ quivers\textquotedblright\ e.g., \cite{Cecotti:2010fi, Cecotti:2011gu, Alim:2011ae,
Alim:2011kw, Cecotti:2013sza}, as well as the method of spectral networks,
e.g. \cite{Gaiotto:2009hg, Gaiotto:2012rg}. Since we have an explicit
presentation for all of the period integrals and we can track the dependence
on moduli, we shall use the method of BPS quivers.

The main idea in the BPS quiver method is to recognize that all of the BPS
particles are obtained as bound states of smaller elementary constituent particles. The
number of independent charges for these particles is completely fixed by the
number of $U(1)$ factors of the model. For each gauged $U(1)$, we get two
charges (one electric and one magnetic), while for each $U(1)$ flavor symmetry
we get one charge (just electric). The dynamics governing the stability of a configuration
is encoded by a supersymmetric quiver quantum mechanics (SQM)
with four conserved supercharges \cite{Denef:2002ru}. The quiver is determined by
the elementary constituents as follows: it has nodes in one to one correspondence with
the charges of the elementary constituents and directed arrows between two
such quiver nodes specified by the Dirac pairing for these charges. In
string theory terms, we view the nodes as candidate BPS\ objects, and the
directed arrows as open strings which stretch from one BPS\ object to the next.
The existence of a bound state of given charge corresponds to the
existence of a ground state for the corresponding SQM \cite{Denef:2002ru, Cecotti:2011rv}.

For the Argyres-Douglas theory, the total number of generators of the Coulomb
branch charge lattice is $2+1=3$. Indeed, we can express the charge
of a candidate state as a three component vector which we write as a linear
combination of the form:%
\begin{equation}
(n_{\text{elec}},n_{\text{mag}},q_{\text{flav}})=\gamma
=\underset{i=1}{\overset{2r+f}{\sum}}n_{i}\gamma_{i}\text{ \ \ for \ \ }%
n_{i}\geq0. \label{lincombo}%
\end{equation}
Here, the $\gamma_{i}$ are the constituent charges out of which all other
stable bound states are constructed.
%For each $\gamma_{i}$ we have a quiver node,
%and the number of directed arrows between a pair of nodes $\gamma_{i}$ and
%$\gamma_{j}$ is given by the Dirac pairing $\langle \gamma_{i},\gamma_{j}\rangle_D$.

Now, as we vary the value of the complex phase in $u$, we can expect some new
bound states to enter or exit the spectrum. For the $H_{1}$ Argyres-Douglas
theory, the full list of candidate states is dictated by the root space of the
corresponding $A_{3}$ lattice \cite{Shapere:1999xr}:%
\begin{equation}
\text{Candidate Charges}=\{\pm\gamma_{1},\pm\gamma_{2},\pm\gamma_{3}%
,\pm(\gamma_{1}+\gamma_{2}),\pm(\gamma_{2}+\gamma_{3}),\pm(\gamma_{1}%
+\gamma_{2}+\gamma_{3})\}. \label{candidates}%
\end{equation}
For this model, all the stable BPS states are hypermultiplets.\footnote{ Borrowing from
standard techniques in soliton theory, the spin of a BPS multiplet is
determined `quantizing' the moduli space of vacua for the SQM
\cite{Witten:1996qb, Denef:2002ru} (see also \cite{DelZotto:2014bga})}

The task of finding the spectrum of stable states therefore decomposes into
two pieces. First, we need to determine a good quiver basis $\{\gamma_{i}\}_{i}$ in
the sense of references \cite{Alim:2011ae,Cecotti:2011rv} and second, we determine which
values of $n_{i}$ in equation (\ref{lincombo}) lead to stable particles.\footnote{ We find, however, that
the technical definition specified in \cite{Alim:2011ae,Cecotti:2011rv} for a good quiver basis is not
enough to determine it uniquely, there is an extra condition (compatibility among mutations and wall-crossings)
which needs to be imposed. The details of this point are discussed in appendix \ref{app:WC_anal}.}
The actual presentation of the quiver as well as the spectrum of
stable particles will depend on the particular region of moduli space where we
are located. The basis of
charges we use to construct our bound states will change, i.e. we have a
transformation of the form:%
\begin{equation}
\gamma_{i}\mapsto\gamma_{i}^{\prime}=\underset{j}{%
%TCIMACRO{\dsum }%
%BeginExpansion
{\displaystyle\sum}
%EndExpansion
}M_{ij}\gamma_{j}, \label{mutato}%
\end{equation}
for $M_{ij}$ an integer valued matrix.\footnote{ The precise form of the allowed matrices $M_{ij}$
is subject to the same caveat discussed in footnote \ref{notsp2r}.} This leads to a \textquotedblleft
mutation\textquotedblright\ or Seiberg duality on the quiver SQM.
The candidate physical charges of line (\ref{candidates}) can also change, i.e., we
build our spectrum of candidates using $\gamma_{i}^{\prime}$ rather than
$\gamma_{i}$. A mutation simply reflects the fact that the structure of
composite objects may change as we change the moduli / parameters of the
model: an object which looks elementary in one frame, can look like a bound state in another.
In string theory terms, this means that we must alter the BPS\ states
used to construct bound states, and correspondingly the spectrum of open
strings will also change. On top of that the actual spectrum of stable BPS states can change as we move
in moduli space (wall-crossing phase transitions).

Since we have an explicit presentation of the various period integrals, it is
straightforward for us to sweep over possible choices of charge assignments.
The main complication is to ensure that we have indeed found all of the stable
particles at a given point in the moduli space, i.e. wall crossing.
%Thankfully, this has been analyzed in great detail elsewhere, so we can
%confine our discussion to general lessons available from working at generic
%points in moduli space.
% NDR: I have commented this sentence because we did this for the first time in this detail in appendix A

At a qualitative level, there are three general regimes of possible interest:%
\begin{align}
\text{Large Mass}  &  \text{: \ \ \ \ }\left\vert m\right\vert \gg\left\vert
u^{3/4}\right\vert \\
\text{Tuned Mass}  &  \text{: \ \ \ }\left(  \frac{u}{3}\right)  ^{3}%
+4m^{4}\simeq0\\
\text{Small Mass}  &  \text{: \ \ \ \ }\left\vert m\right\vert \ll\left\vert
u^{3/4}\right\vert .
\end{align}
For illustrative purposes, we study in detail the large mass regime. We shall
also explain how a similar analysis applies at small mass parameters.

Consider, then, the large mass regime. Here, we have $\tau_{\text{extra}%
}\simeq e^{2\pi i/6}\simeq e^{2\pi i/3}$, so we are at strong coupling. An
additional simplification is that we always expect the lightest object to be
neutral under the flavor symmetry. To determine the spectrum near this point,
it is helpful to rely on the existing analysis of BPS\ quivers presented for
example in \cite{Alim:2011ae, Cecotti:2013sza}. For the $H_1$ theory,
there are always at least three stable BPS states corresponding to the three nodes of the BPS\ quiver.
These are always ${\mathcal N}=2$ hypermultiplets. In addition to these three states, there can in principle be others which are also stable.

\begin{figure}[ptb]
\centering
\begin{tabular}{cc}
\includegraphics[
scale = 0.7, trim = 0mm 0mm 0mm 0mm
]{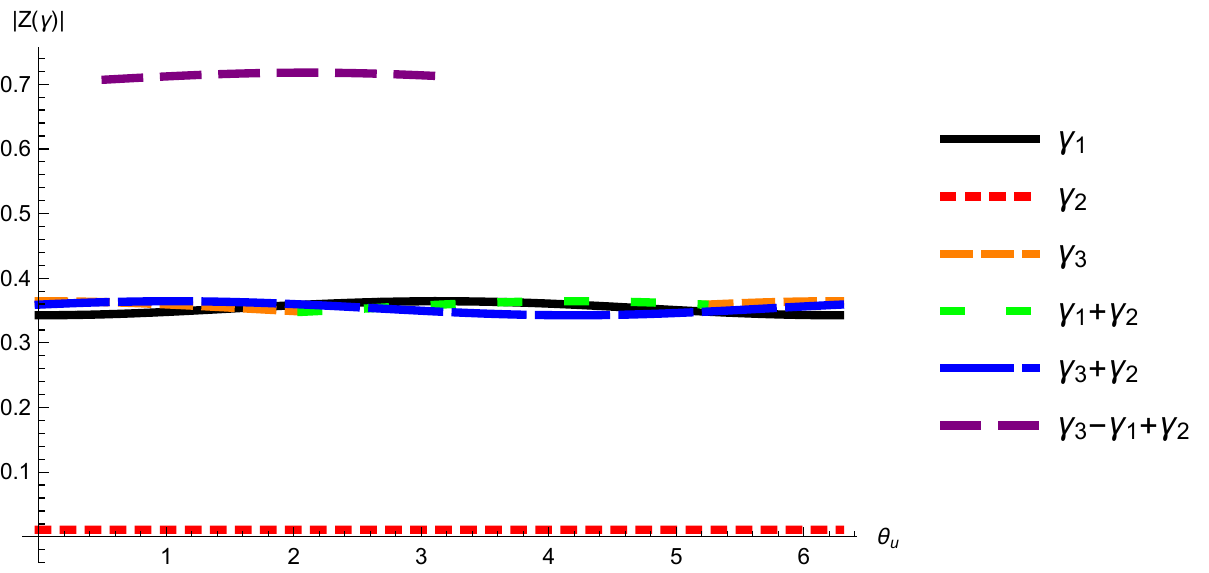} &
\includegraphics[
scale = 0.7, trim = 0mm 0mm 0mm 0mm
]{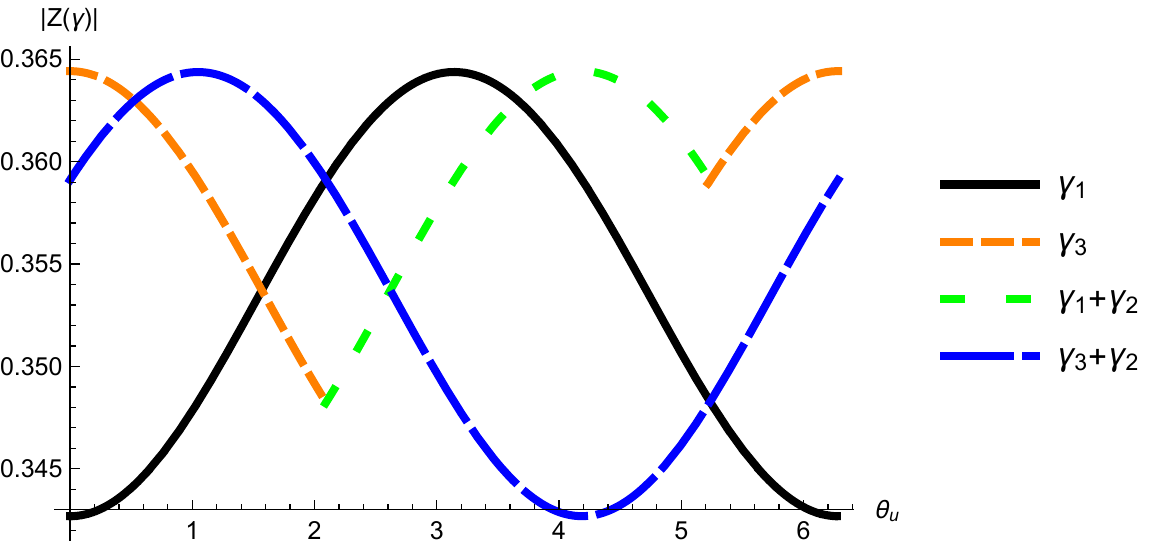}\\
\end{tabular}
\caption{\textsc{left}: Plot of the spectrum of masses for the stable states
as a function of $\theta$ in the large mass regime.
For numerical purposes, we take $u = 0.1 \exp(i \theta)$ and $m = 1$. Notice that a BPS state with charge $\gamma_3 -\gamma_1 + \gamma_2$ enters the spectrum in the region $0.525 < \theta < 3.65$. This is possible precisely because the BPS quiver relevant in that region is a mutant of the one in line \eqref{goodquiver} (see appendix \ref{app:WC_anal} for the details). \textsc{right}: Magnified region of the plot of $\vert Z \vert$ with $u = 0.1 \exp(i \theta)$ and $m = 1$, which shows $\gamma_3$ destabilizing and $\gamma_1 + \gamma_2$ stabilizing in complementary regions.}%
\label{largemass}%
\end{figure}

We find that when $\left\vert u\right\vert =0.1$ and $m=1$, we are
effectively in the large mass regime. So let us turn to an analysis of the BPS
quiver in this regime. To illustrate, suppose that we hold fixed the
parameters:%
\begin{equation}\label{illustrativelargemass}
u=0.1\exp(i\theta)\text{, \ \ }m=1\text{.}%
\end{equation}
When $ 5\pi/3  \leq\theta\leq 2\pi$, the quiver SQM governing the dynamics of the BPS solitons is the quiver
with nodes:%
\begin{equation}\label{goodquiver}
\text{BPS\ Quiver: \ \ \ }\gamma_{1}\longrightarrow \gamma_{2} \longleftarrow
\gamma_{3},
\end{equation}
with:%
\begin{equation}%
\begin{tabular}
[c]{|c|c|c|c|}\hline
node
%TCIMACRO{\TEXTsymbol{\backslash} }%
%BeginExpansion
$\backslash$
%EndExpansion
charge & $n_{\text{elec}}$ & $n_{\text{mag}}$ & $q_{\text{flav}}$\\\hline
$\gamma_{1}$ & $+1$ & $+1$ & $+1/2$\\\hline
$\gamma_{2}$ & $-1$ & $0$ & $0$\\\hline
$\gamma_{3}$ & $+1$ & $+1$ & $-1/2$\\\hline
\end{tabular}
\ \ \ \ \ ,
\end{equation}
so the 3 hypermultiplets with charges $\gamma_1$, $\gamma_2$, and $\gamma_3$ are
the elementary BPS states in this region of moduli space. Notice that we have a stable \textquotedblleft dark
electron\textquotedblright\ with charge $\gamma_{2}$. This is a stable BPS particle which is neutral under the flavor symmetry. For these values of $\theta$ there is an additional bound state with charge $\gamma_2 + \gamma_3$ in the spectrum. Now, as we vary the phase
$\theta$, we can expect that some of these objects ceases to be elementary
and decay to other stable constituents. To figure out the possible changes as
we move around, we need to explore the various mutants quiver
SQMs occurring as we vary $\theta$ in line \eqref{illustrativelargemass}.
The actual pattern of wall-crossings is analyzed in details in appendix \ref{app:WC_anal}.
The precise structure of the BPS spectrum as a function of $\theta$ is plotted in figure \ref{largemass}.
We find that in sweeping over all values of the phases for $|u|\sim 0.1$ the ``dark
electron'' with charge $\gamma_{2}$
remains a stable object of the spectrum.

\begin{figure}[ptb]
\centering
\begin{tabular}{cc}
\includegraphics[
scale = 0.7, trim = 0mm 0mm 0mm 0mm
]{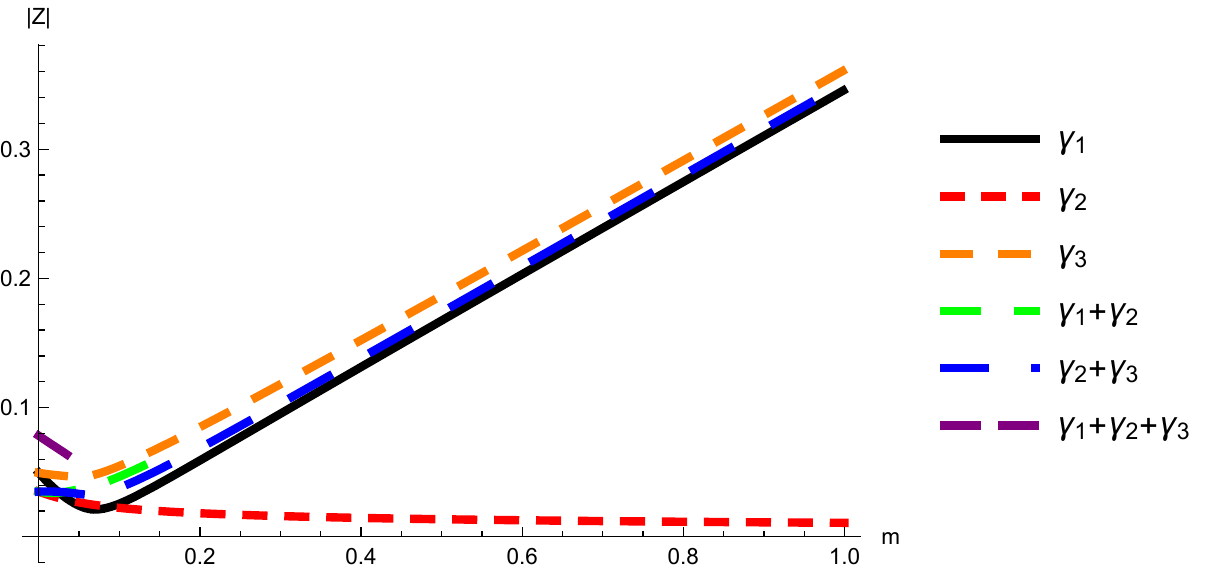} &
\includegraphics[
scale = 0.7, trim = 0mm 0mm 0mm 0mm
]{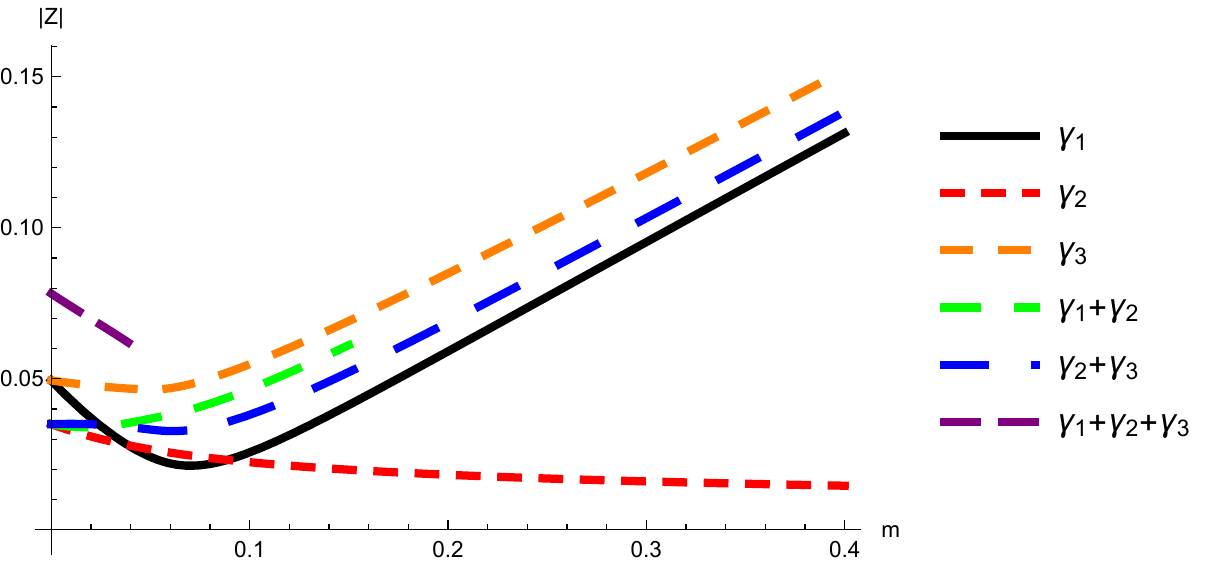}\\
\end{tabular}
\caption{\textsc{left}: Plot of the spectrum of masses for the stable states
as a function of $m$ interpolating from the large mass regime $m=1$ to $m=0$.
For numerical purposes, we take $u = 0.1 \exp(i 5.5)$. The whole deformation is covered by the BPS quiver in line \eqref{goodquiver}: no mutation occurs. \textsc{right}: Magnified region of the plot which shows 1.) the region where the dark dyon stabilizes and 2.) the small region where the dark electron ceases to be the lightest massive excitation.}%
\label{smalltolargemass}%
\end{figure}

Let us also note that although the \textquotedblleft dark
dyon\textquotedblright\ with charge $\gamma_{1}+\gamma_{2}+\gamma_{3}$ has
lower mass than its flavor charged counterparts, it is nevertheless not a
stable object in the large mass regime for $|u| \sim 0.1$. Rather, it can enter the
spectrum as we decrease the value of $m$ (see figure \ref{smalltolargemass} as well as appendix \ref{app:WC_anal}).

Similar analyses can be carried out for all of the regions of moduli space and
mass parameter space. An important point is that near the region $m=0$, we
also have a restored $SU(2)$ flavor symmetry, so as a consequence, the states
have a mass degeneracy compatible with this fact. Another interesting feature
close to this region is that the state of charge $\gamma_{2}$ is not always
the lightest in the spectrum.

\subsubsection{The Dark Electron and Dark Dyon}

From our analysis of the mass spectrum of the $H_{1}$ Argyres-Douglas theory,
we can also draw some conclusions about the spectrum of stable particles which
are neutral under the visible sector gauge coupling. For both the large and
small mass regime, the states of charge $\pm\gamma_{2}$ are stable. We refer
to this as a \textquotedblleft dark electron\textquotedblright\ since it only
has electric charge under the extra sector $U(1)$. Additionally, in some
regions of parameter space, there is another flavor neutral state which in a
suitable basis of electric and magnetic charges has charge vector $\pm
(\gamma_{1}+\gamma_{2}+\gamma_{3})$ which we refer to as the \textquotedblleft
dark dyon\textquotedblright\ since it has both electric and magnetic charge
under the extra sector $U(1)$.

Now, even when these states are unstable, they can still play an important
role in scattering events between the visible and hidden sector. The reason is
that with a sufficiently energetic process in the extra sector, we may still
be able to generate such charged states. Since we can also calculate the
effects of kinetic mixing, we now ask what the effective electric charge under
the visible sector $U(1)$ is for each of these states. The effective electric
charge follows from our formula for dark Rutherford scattering presented in
equation (\ref{qeffdef}). To keep the analysis simple yet tractable, we shall
primarily focus on the single slice of parameters $u=0.1$, with $m$ varying by
powers of $5$. As in our earlier analyses, we work in dimensionless units,
i.e., depending on the scale of conformal symmetry breaking (dictated by its
coupling to other sectors) the actual mass of the state could be anywhere from
the TeV scale to the GUT\ scale. Here then, is the list of effective electric
charges as we vary the value of $m$:%
\begin{align}
&  \text{%
\begin{tabular}
[c]{|l|l|l|l|l|}\hline
Dark Electron & $m=0.04$ & $m=0.2$ & $m=1.0$ & $m=5.0$\\\hline
$\left\vert q_{\text{eff}}\left(  \gamma_{2}\right)  \right\vert $ &
$2.0\times10^{-2}$ & $1.0\times10^{-3}$ & $1.3\times10^{-5}$ & $1.8\times
10^{-7}$\\\hline
$\left\vert Z(\gamma_{2})\right\vert $ & $2.9\times10^{-2}$ & $1.8\times
10^{-2}$ & $1.1\times10^{-2}$ & $6.3\times10^{-3}$\\\hline
\end{tabular}
,}\label{darkone}\\
&  \text{%
\begin{tabular}
[c]{|l|l|l|l|l|}\hline
Dark Dyon & $m=0.04$ & $m=0.2$ & $m=1.0$ & $m=5.0$\\\hline
$\left\vert q_{\text{eff}}\left(  \gamma_{1}+\gamma_{2}+\gamma_{3}\right)
\right\vert $ & $4.0\times10^{-1}$ & $8.1\times10^{-2}$ & $1.0\times10^{-2}$ &
$1.2\times10^{-3}$\\\hline
$\left\vert Z\left(  \gamma_{1}+\gamma_{2}+\gamma_{3}\right)  \right\vert $ &
$5.6\times10^{-2}$ & $3.2\times10^{-2}$ & $1.9\times10^{-2}$ & $1.1\times
10^{-2}$\\\hline
\end{tabular}
}\label{darktwo}%
\end{align}
where for reference we have also included the corresponding values of the
central charge. Again, we emphasize that the dark dyon is not stable in some
regions of parameter space, e.g., in the large mass regime $\left\vert
m\right\vert \gtrsim1$.

\section{Phenomenological Toy Models \label{sec:TOY}}

Having spelled out the main technical elements of how to compute kinetic
mixing at strong coupling, we now turn to some aspects of how these models
embed in more realistic phenomenological scenarios. Even so, we will keep our
discussion at the level of toy models, using the $H_{1}$ Argyres-Douglas
theory as our primary example.

Indeed, in the context of string constructions, the Argyres-Douglas theory
should be viewed as a subsector of a more complete model. From a bottom up
perspective, however, we can view deformations of the $H_{1}$ Argyres-Douglas
theory as a candidate extra sector in its own right. Even in this case,
however, there are several moving parts which can impact the resulting phenomenology.

The rest of this section is organized as follows. First, we place the $H_{1}$
Argyres-Douglas theory in the context of more general stringy constructions
which incorporate the Standard Model. After this, we explain how different
scales of conformal symmetry breaking lead to different types of
phenomenological scenarios.

\subsection{String-Motivated Examples}

One of the motivations for this work is the fact that string constructions
typically contain extra $U(1)$'s which can mix with the visible sector $U(1)$.
To illustrate the general suite of ideas, we focus on the class of extra
sectors introduced in \cite{Heckman:2010fh, Heckman:2011sw, Heckman:2010qv,
Heckman:2011hu, Cecotti:2010bp}. In these models, the Standard Model is
realized from a stack of intersecting seven-branes, and the extra sector is
realized by a probe D3-brane. This D3-brane is energetically attracted to the
visible sector by the same mechanism which generates quark and lepton masses
and mixing angles \cite{Cecotti:2009zf, Heckman:2008qa} (see also
\cite{Martucci:2006ij, Marchesano:2009rz}). As a passing remark, we note that
in constructions of the Standard Model via heterotic M-theory, a similar class
of extra sectors are realized by M5-branes wrapped on a curve of the
compactification manifold.

A priori, there may be other local minima for the D3-brane, so fluxes may
localize it at other points of the compactification manifold. Indeed, we can
expect there to typically be many such D3-branes. The total number in a
general type IIB\ background is given by the formula \cite{Sethi:1996es}:
\begin{equation}
N_{D3}=\frac{\chi(CY_{4})}{24}+\underset{B}{\int}H_{NS}\wedge H_{RR},
\label{D3branes}%
\end{equation}
where $\chi(CY_{4})$ is the Euler characteristic of the elliptically fibered
Calabi-Yau fourfold used to define an F-theory background, and $H_{NS}$ and
$H_{RR}$ are three-form fluxes which are integrated over the six-dimensional
internal spacetime $B$. Values of $\chi(CY_{4})/24$ can range from $O(10^{2})$
to $O(10^{4})$ (see e.g., \cite{Klemm:1996ts, Denef:2008wq}), so depending on
the choice of background fluxes, one can contemplate scenarios with either
many D3-branes, or only a small number.

One of the interesting features of kinetic mixing is that because it comes
from integrating out heavy states to generate marginal couplings, we can
expect there to be possible contributions to electric and magnetic kinetic
mixing even for those D3-branes which are far removed from our visible sector
stack. So, even for extra sector models where other direct couplings to the
Standard Model are suppressed (as they typically will be), kinetic mixing at
strong coupling can still survive.

Let us now turn to more details of the resulting effective field theory on a
D3-brane. In the limit where the D3-brane is close to the Standard Model stack
of intersecting seven-branes, we can visualize this extra sector as an
$\mathcal{N}=2$ superconformal field theory with $E_{8}$ flavor symmetry
\cite{Minahan:1996fg, Minahan:1996cj} which is subject to $\mathcal{N}=1$
relevant and marginal deformations which induce a flow to an $\mathcal{N}=1$
superconformal field theory in which the flavor symmetry of this IR\ theory
includes the gauge group of the Standard Model \cite{Heckman:2010fh,
Heckman:2011sw, Heckman:2010qv, Heckman:2011hu, Cecotti:2010bp}.

We organize our discussion according to the decomposition of $SU(5)_{\text{GUT}}\times SU(5)_{\bot} \subset E_{8}$,
with corresponding mass deformations valued in the adjoint representations, i.e. we schematically
write $\phi_{\text{GUT}}$ and $\phi_{\bot}$ for these Coulomb branch
parameters. Geometrically, the main idea is that the Coulomb branch parameter
$u$ describes the position of a D3-brane normal to the $SU(5)_{\text{GUT}}$
seven-brane. There are also two complex directions $u_{1}$ and $u_{2}$
parallel to the seven-brane. In the associated field theory, $u_{1}\oplus
u_{2}$ parameterize a decoupled hypermultiplet. To get an $\mathcal{N}=1$
deformation, we therefore allow the mass parameters of the theory to depend on
$u_{1}$ and $u_{2}$, so we make the substitution $\phi_{\bot}\mapsto\phi
_{\bot}(u_{1},u_{2})$. Additionally, we need not require that $\phi_{\bot}$ is
even diagonal. We can also consider mass deformations which break
$SU(5)_{\text{GUT}}$ to $SU(3)\times SU(2)\times U(1)$, i.e. by taking a mass
deformation in the same direction as $U(1)_{Y}$.

To apply the methods of the present paper, we must also assume that the
deformation to an $\mathcal{N}=1$ vacuum is sufficiently mild, i.e. we have a
"short flow" from a neighboring $\mathcal{N}=2$ theory. Now, even though we
only have $\mathcal{N}=1$ supersymmetry, there is still a notion of a
Seiberg-Witten curve, with the Seiberg-Witten differential now replaced by a
meromorphic four-form of a non-compact Calabi-Yau fourfold. The main caveat to
extracting numerical estimates, however, is that the physical couplings may
now receive non-trivial contributions from wave function renormalization. This
shows up quite directly in other contexts as corrections to the scaling
dimensions of operators in the deformed theory, see e.g.,
\cite{Heckman:2011sw, Heckman:2011hu}.

While we leave a complete analysis of this more involved case to future work,
it is interesting to already explore some of the general features of these
models. First of all, we see that if we take most of the mass parameters to be
of the GUT\ scale or higher, then the lightest states which can meaningfully
participate at low energies will be those which are neutral under the flavor
symmetries. As we have already seen in the $H_{1}$ Argyres-Douglas theory,
there is a lightest state which is neutral under all such flavor symmetries,
with mass controlled primarily by the Coulomb branch parameter. Additionally,
we can re-incorporate some of the effects of heavier states of the model.
These will show up as line operators (that is, heavy quarks) of the theory,
and we can also contemplate bound states of comparatively light objects to
these line operators. The excitation scale for these heavy objects can
naturally be at the GUT\ scale or higher, so in this sense, their direct
relevance for phenomenology may be more limited. It is interesting to note,
however, that in some cases we can tune parameters of the string-based model
to realize excitations of these objects at lower energy scales. Indeed, an
intriguing novelty of rank one theories with larger flavor symmetry groups
such as $E_{6}$, $E_{7}$ and $E_{8}$ is the presence of whole Regge
trajectories of stable objects in certain ranges of moduli space
\cite{Cecotti:2013sza, Cordova:2015vma, Cordova:2015zra, Hollands:2016kgm}.
This clearly leads to a rich class of possibilities, which would be quite
interesting to study in future work.

\subsection{Mass Scales}

To make more contact with model building considerations, we clearly need to
specify possible mass scales for our model. Since we have an extra sector with
approximate conformal symmetry, we expect that the masses of the extra sector
states will be dictated by the scale of conformal symmetry breaking. Even in
this case, however, we can get different mass hierarchies, since as we saw in
the case of the $H_{1}$ Argyres-Douglas theory, taking the mass parameter $m$
very large still leaves us with a light state which we referred to as the
\textquotedblleft dark electron.\textquotedblright\ In other regimes of
parameter space, this can also be accompanied by a \textquotedblleft dark
dyon.\textquotedblright\ Let us step through the different kinds of scenarios
associated with each sort of mass scale.

\subsubsection{GUT Scale Masses}

Suppose we take the simplest scenario in which all hypermultiplets have GUT
scale masses. This possibility is also well-motivated in the context of string
constructions. In this case, we expect to be left at low energies with a
collection of $U(1)$ gauge bosons and their $\mathcal{N}=2$ superpartners.
Transmission of supersymmetry breaking to the extra sector will then lead to
further mass splittings amongst the states.

The phenomenological bounds on extra decoupled $U(1)$'s are quite weak, since
without any charged states from the extra sector, there is no way to directly
detect these vector bosons.

The caveat to this statement is that we also have the $\mathcal{N}=2$
superpartners, which include a gaugino, and a decoupled $\mathcal{N}=1$ chiral
multiplet. As we have already remarked, the extra sector gauginos can mix with
visible sector gauginos. These mixing terms depend on the details of
supersymmetry breaking, but we have shown in section \ref{sec:SUSY} how to
calculate these contributions in certain supersymmetry breaking scenarios by
computing the second derivatives of $a^{D}$ with respect to $a$ and $m$. Some
aspects of the phenomenology of these photini mixing have been studied for
example, in reference \cite{Arvanitaki:2009hb}.

Consider next the $\mathcal{N}=1$ chiral multiplet. In the limit of exact
$\mathcal{N}=2$ supersymmetry, the presence of a complex scalar with no
potential suggests the presence of a modulus, which if left unstabilized, can
lead to a cosmological history in which the energy density is dominated by
such a rolling scalar.

There is a simple way to ameliorate this issue by introducing an overall
superpotential deformation of the system, i.e. $W(u)$, for the Coulomb branch
parameter. For us to continue to use our $\mathcal{N}=2$ supersymmetric
approximation, we simply need to require that the mass scale for the scalar is
small compared to those of the charged states, i.e. that $m_{a}\ll\langle a
\rangle$, with $a$ the local expression for the Coulomb branch scalar. This is
technically natural since such mass terms are conformally suppressed in this
class of models \cite{Heckman:2011sw, Heckman:2010qv}.

\subsubsection{TeV and Sub-TeV Scale Masses}

It is also natural to consider scenarios in which some of the extra sector
states have masses far below the GUT scale. For example, if the D3-brane
remains close to the Standard Model stack, we can still expect some flavor
neutral hypermultiplets to survive to much lower energies. Again, this is
technically natural since a superpotential deformation for the Coulomb branch
parameter can be conformally suppressed \cite{Heckman:2011sw, Heckman:2010qv}.
In such cases, transmission of supersymmetry breaking to the extra sector will
also contribute to the masses of these states. We can also see from our
analysis near lines (\ref{darkone}) and (\ref{darktwo}) that the effective
electric charge for these flavor neutral states can be quite small. For some
discussion on cosmological constraints on millicharged particles, as well as
scenarios with an exactly massless $U(1)$ decoupled from the Standard Model,
see respectively \cite{Davidson:2000hf} and \cite{Ackerman:mha} (see also \cite{Foot:2014uba}).

In the TeV scale mass range, much of the phenomenology is dictated by whether
the extra sector $U(1)$ is electrically screened / magnetically confined or
remains as a long range force carrier. Some aspects of the former case were
studied in detail in reference \cite{Heckman:2011sw} to which we refer the
interested reader for further details. In this case, we get string-motivated
examples of asymmetric dark matter models with order $10$ GeV masses for dark
matter. The sub-TeV mass scale originates from a seesaw like mechanism for
dark states connected with partial breaking to $\mathcal{N}=1$ supersymmetry
\cite{Heckman:2011sw}. Even lower mass scales are potentially possible, though
the presence of heavier extra sector states charged under the visible sector
means that we must exercise some care in building such models.

If, on the other hand, we assume that the extra sector $U(1)$ remains as a
long range force carrier, then we have a conserved electric and magnetic
charge, and so we can also expect there to be stable dark states. We have also
seen that visible sector charged states can be decoupled.

Assuming we have a TeV scale dark state, we can estimate its cosmological
relic abundance. The fact that we have kinetic mixing with the visible sector,
as well as a strongly coupled extra sector means that the overall thermally
produced relic abundance will be lower than that of the standard WIMP example.
For example, letting $\Omega_{\text{extra}}$ denote the relic abundance of
such an extra sector state, and $\Omega_{\text{DM}}$ that of WIMP\ dark
matter, we have:
\begin{equation}
\frac{\Omega_{\text{extra}}}{\Omega_{\text{DM}}}\sim\frac{\alpha_{\text{WIMP}%
}^{2}}{\alpha_{\text{extra}}^{2}}\frac{M_{\text{extra}}^{2}}{M_{\text{WIMP}%
}^{2}}\sim\left(  \frac{10^{-3}}{\alpha_{\text{extra}}^{2}}\right)  \left(
\frac{M_{\text{extra}}}{1\text{ TeV}}\right)  ^{2},
\end{equation}
so we see that if our extra sector states are around the TeV scale, an order
one value for $\alpha_{\text{extra}}$ suppresses the overall contribution of
these states. For this reason, we see that any individual extra sector will
make only a small contribution to the net relic abundance, i.e. we can easily
satisfy various cosmological bounds. Observe also that if we have a
non-thermal epoch in the evolution of the Universe, i.e., one with a late
decaying scalar, it can also be beneficial to overproduce this relic
abundance, as is common in some string based constructions
\cite{Heckman:2008jy, Acharya:2009zt}.

Aside from their potential role in cosmology (if we have multiple decoupled
extra sectors to obtain a suitable relic abundance), we now have the strongly
coupled analogue of extra charged states which could be generated in collider
experiments. Indeed, we have also explained how these extra sector states can
produce an effective electric charge (c.f. equation (\ref{qeffdef})). This
leads to generalizations of the standard $Z^{\prime}$ scenario which it would
be interesting to study further. It is important to emphasize, however, that
the strongly coupled nature of the extra sector means that some of the
implicit assumptions usually made in the analysis of $Z^{\prime}$ models
should be revisited before drawing any definite conclusions on this class of
models. We leave a full analysis of this possibility for future work.

\section{Conclusions \label{sec:CONC}}

Kinetic mixing at strong coupling is well-motivated from both a top down and
bottom up perspective. We have shown how to extract the leading order mixing
terms for an extra sector with approximate $\mathcal{N}=2$ supersymmetry, and
commented on their potential role in phenomenological scenarios. In the
remainder of this section we discuss some avenues of future investigation.

It would be interesting to extend our analysis to larger unbroken flavor
symmetry groups for the extra sector. In particular, theories with exceptional
flavor symmetry have a rich spectrum of BPS objects which can also figure in
model building considerations.

A related question is how to carry over our results to models in which
$\mathcal{N}=2$ supersymmetry is broken to $\mathcal{N}=1$ or $\mathcal{N}=0$
supersymmetry. Provided these supersymmetry breaking effects are sufficiently
mild, we anticipate that the formal techniques developed here should be more
broadly applicable.

Finally, it is tempting to speculate that because our $\mathcal{N}=2$ sector
contains a scalar modulus with a flat potential, that this mode could play the
role of an inflaton in slow roll inflation \cite{Heckman:2010fh}, with
reheating triggered by reaching the origin of moduli space. This suggests yet
another potential role for such extra sectors.

\newpage

\section*{Acknowledgements}

We thank C. Cordova, T.T. Dumitrescu, A. Erickcek, P. Langacker, J. Ruderman,
Y. Tachikawa, S.\ Watson, N. Weiner and I. Yavin for helpful discussions. We
also thank J. Ruderman and S. Watson for comments on an earlier draft.
JJH\ thanks the theory groups at Columbia University, the ITS at the
CUNY\ graduate center, and the CCPP at NYU for hospitality during the
completion of this work. The work of MDZ is supported by NSF grant
PHY-1067976. JJH and AM\ are supported by NSF CAREER grant
PHY-1452037. JJH and AM also acknowledge support from the Bahnson Fund at UNC
Chapel Hill as well as the R.~J. Reynolds Industries, Inc. Junior Faculty
Development Award from the Office of the Executive Vice Chancellor and Provost
at UNC Chapel Hill. The work of BW was supported in part by the Science and
Technology Facilities Council Consolidated Grant ST/L000415/1
\textquotedblleft String theory, gauge theory \& duality.\textquotedblright%
\ MDZ, JJH, PK and BW thank the 2013 Simons Center for Geometry and Physics
Summer workshop for hospitality during part of this work. MDZ and JJH also
thank the 2016 Simons Center for Geometry and Physics Summer workshop for
hospitality during the completion of this work. MDZ also thanks the organizers
of the 2016 Amsterdam String Theory workshop for hospitality while this work
was completed.

%\newpage

\appendix

\section{Details on the BPS Spectrum of the $H_1$ Model}\label{app:WC_anal}
\begin{figure}
\begin{tabular}{ccc}
\includegraphics[scale = 0.6]{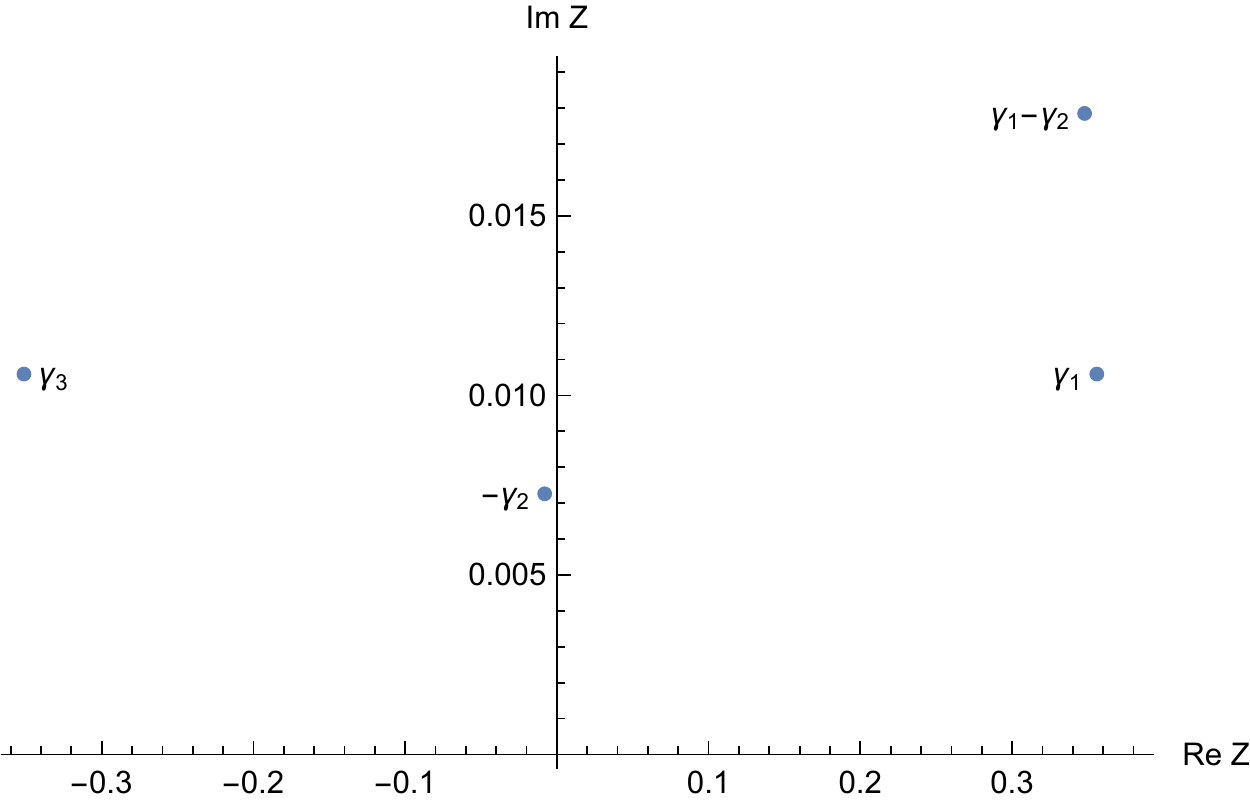}& &
\includegraphics[scale = 0.6]{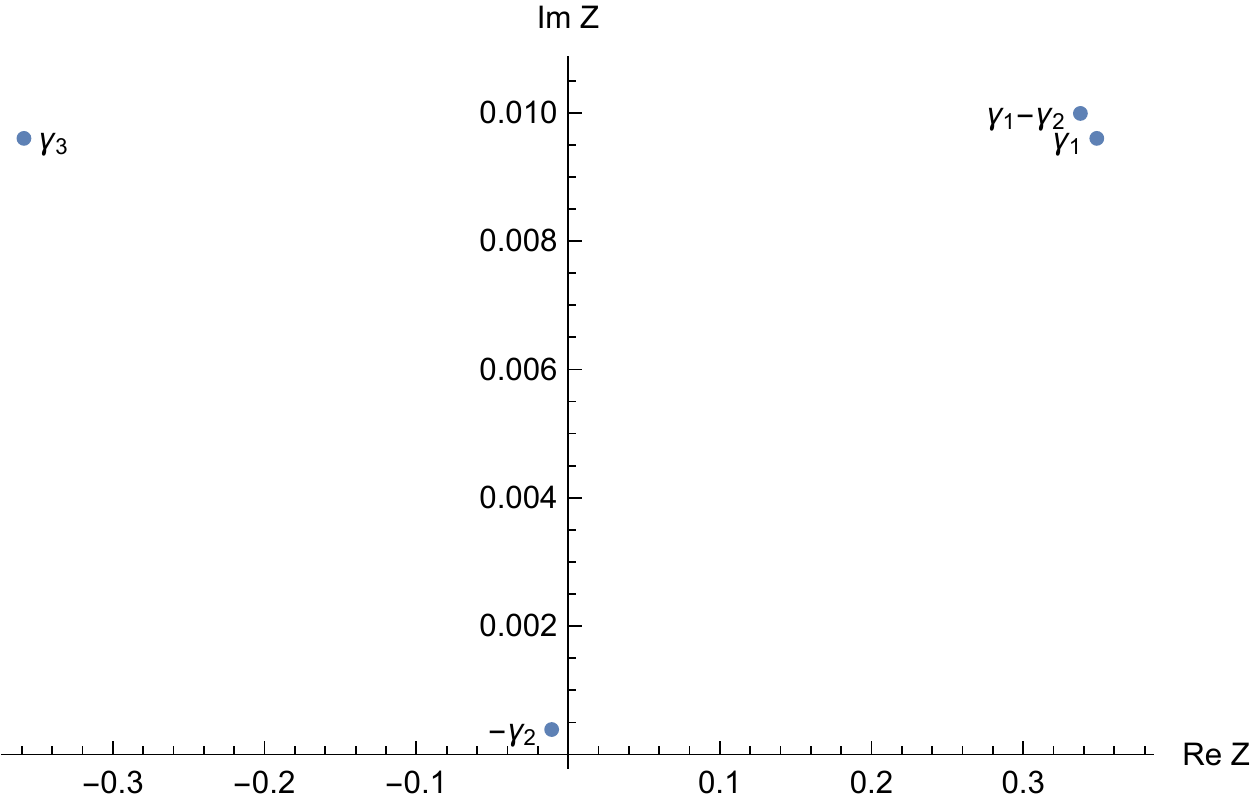}\\
\end{tabular}
\caption{\textsc{left:} Putative quiver basis $\gamma_1,-\gamma_2,\gamma_3$ for $\pi < \theta < 5 \pi/3$.
\textsc{right:} Same basis as $\theta $ approaches $5 \pi/3$ from the left:
one can see that the boundstate $\gamma_1 - \gamma_2$ destabilizes by wall-crossing,
which is in contradiction with the mutation rule for the basis elements as $Z(-\gamma_2)$
exits the upper $Z$-plane from the negative real axis.}\label{WXwrong}
\end{figure}

The BPS quiver computation of the BPS spectrum is achieved by means of quiver representation theory \cite{Denef:2002ru}.
By a standard geometric invariant theory argument, the Higgs branch moduli space of the quiver SQM corresponding to
a state of charge $\gamma$ is equivalent to the moduli space of stable representations of
dimension vector $(N_1,...,N_n)$, where $\gamma = \sum_i N_i \gamma_i$, and $\gamma_i$ is
a good quiver basis in the sense of references \cite{Alim:2011ae,Cecotti:2011rv} (i.e. one for which
1.) the coefficients $N_i$ are either all non-negative integers, or non-positive ones and
2.) $Im Z(\gamma_i) >0$ $\forall$ \, $i=1,...,n$). In our case we find two candidates of good
quiver basis for the $H_1$ model at $u\sim 0.1$ and $m\sim 1$:
the one outlined in the main body of the text and the one given by
$\gamma_1,(-\gamma_2),\gamma_3$ in the region $\pi < \theta < 5 \pi/3$.
Consider the former. See the LHS of figure \ref{WXwrong} to see that indeed it meets requirements 1.) and 2.)
of \cite{Alim:2011ae,Cecotti:2011rv}. The candidate basis $\gamma_1,(-\gamma_2),\gamma_3$ does not
mutate to $\gamma_1,\gamma_2,\gamma_3$ at $\theta=5 \pi/3$, but rather it mutates to
$\gamma_1-\gamma_2,\gamma_2,\gamma_3-\gamma_2$, which leads to an inconsistency.
The quiver  for this putative basis would be
\begin{equation}\gamma_1 \longleftarrow (-\gamma_2) \longrightarrow \gamma_3,\label{wrongone}\end{equation}
from which we see that the representation with dimension vector $(1,1,0)$ is indeed stable.
As $\theta $ approaches $5 \pi/3$ from the left, one can see from figure \ref{WXwrong} that $Z(-\gamma_2)$ exits the upper $Z$-plane from the
negative real axis, which triggers the quiver mutation from the quiver in equation \eqref{wrongone} to
\begin{equation}\gamma_1-\gamma_2 \longrightarrow \gamma_2 \longleftarrow \gamma_3-\gamma_2.\end{equation}
At the same time the state $\gamma_1-\gamma_2$ is wall-crossing away (getting unstable
and disappearing from the spectrum), which is inconsistent with the charges on the nodes of the
mutated quiver, because quiver nodes always correspond to stable particles. This rules out the candidate basis
$\gamma_1,(-\gamma_2),\gamma_3$ with respect to the one we use in the main body of the text, which does not lead to such inconsistencies.

Let us proceed by reviewing the computation of the BPS spectra we summarized
in figures \ref{largemass} and \ref{smalltolargemass}. Let us first consider the large mass regime with $u=0.1$ and $m=e^{i\theta}$.

\begin{figure}
\begin{tabular}{ccc}
\includegraphics[scale = 0.6]{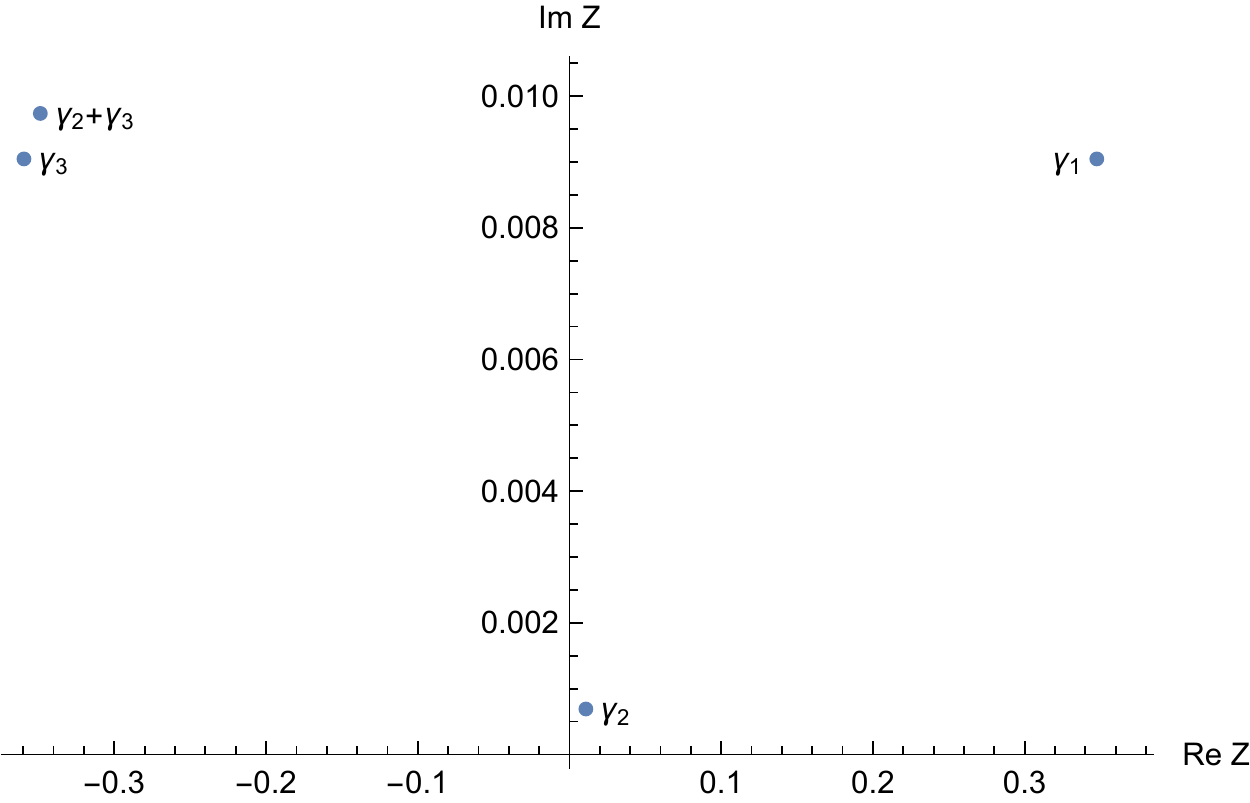}& &
\includegraphics[scale = 0.6]{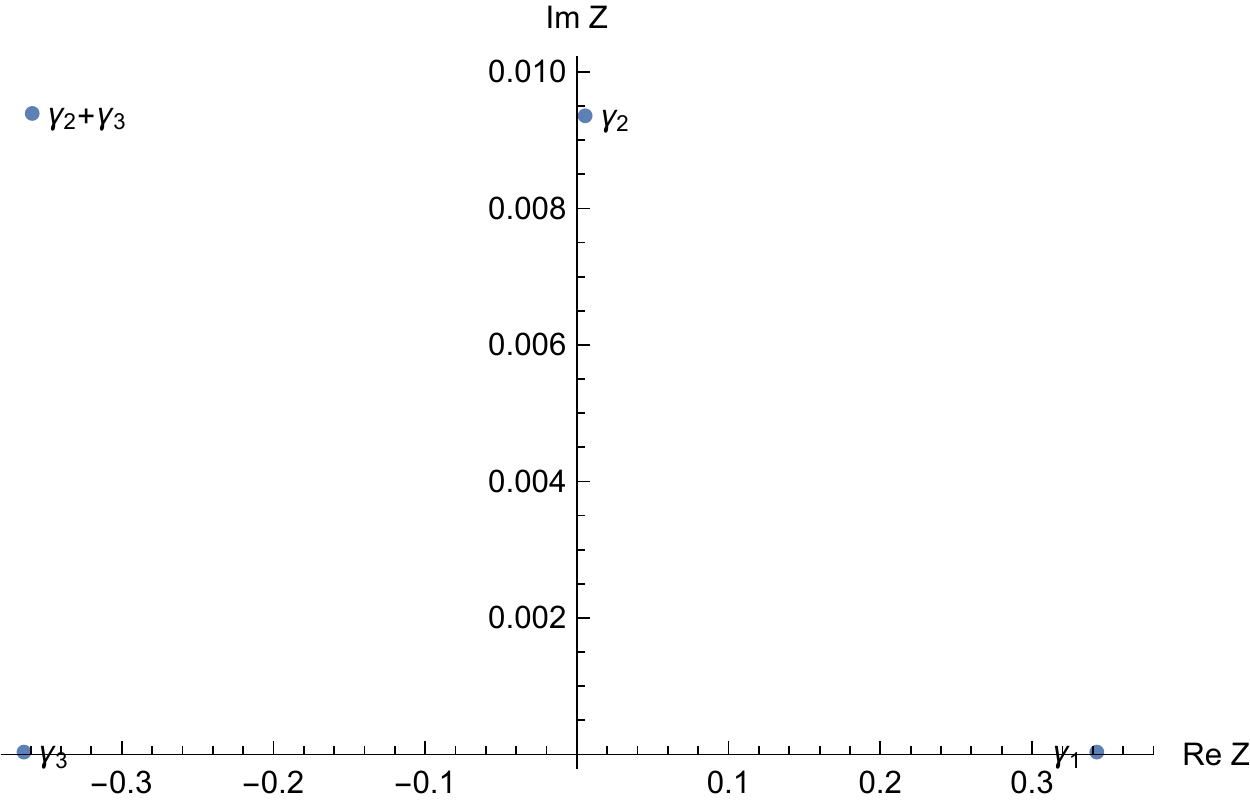}\\\\
\includegraphics[scale = 0.6]{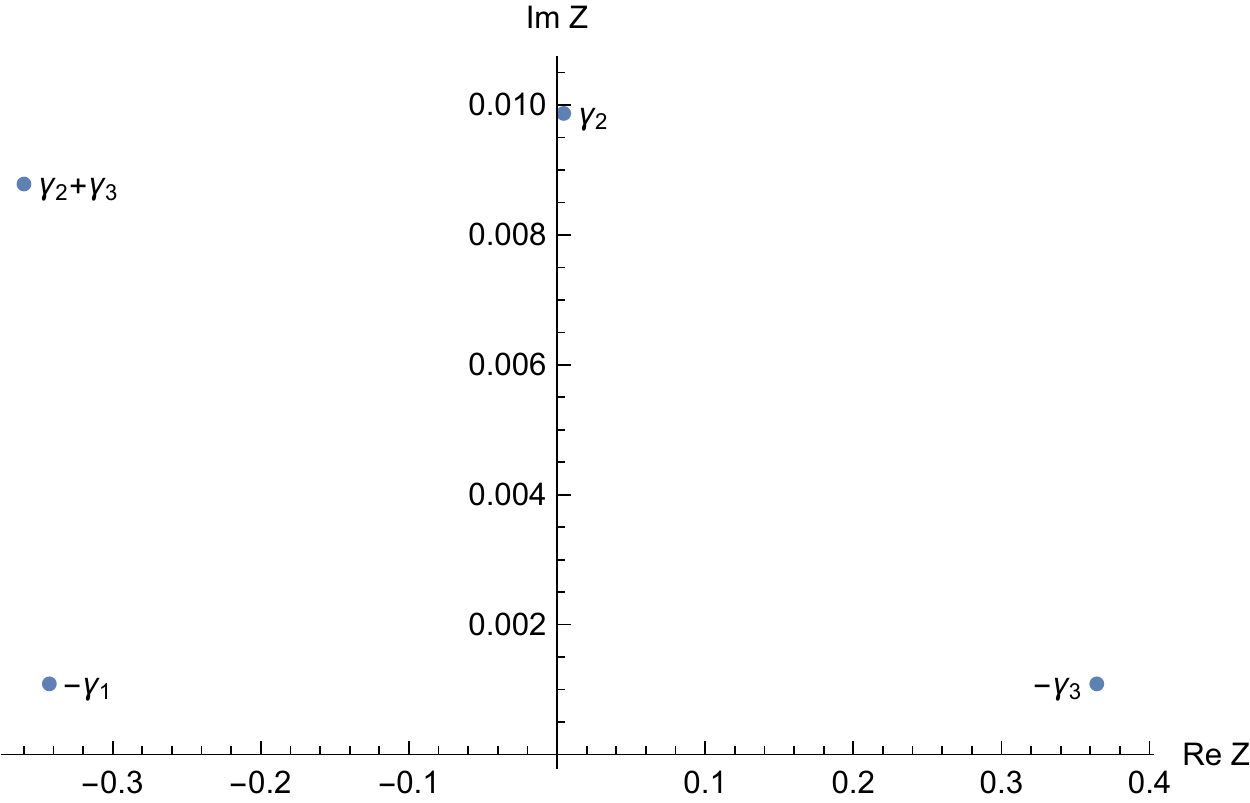}& &
\includegraphics[scale = 0.6]{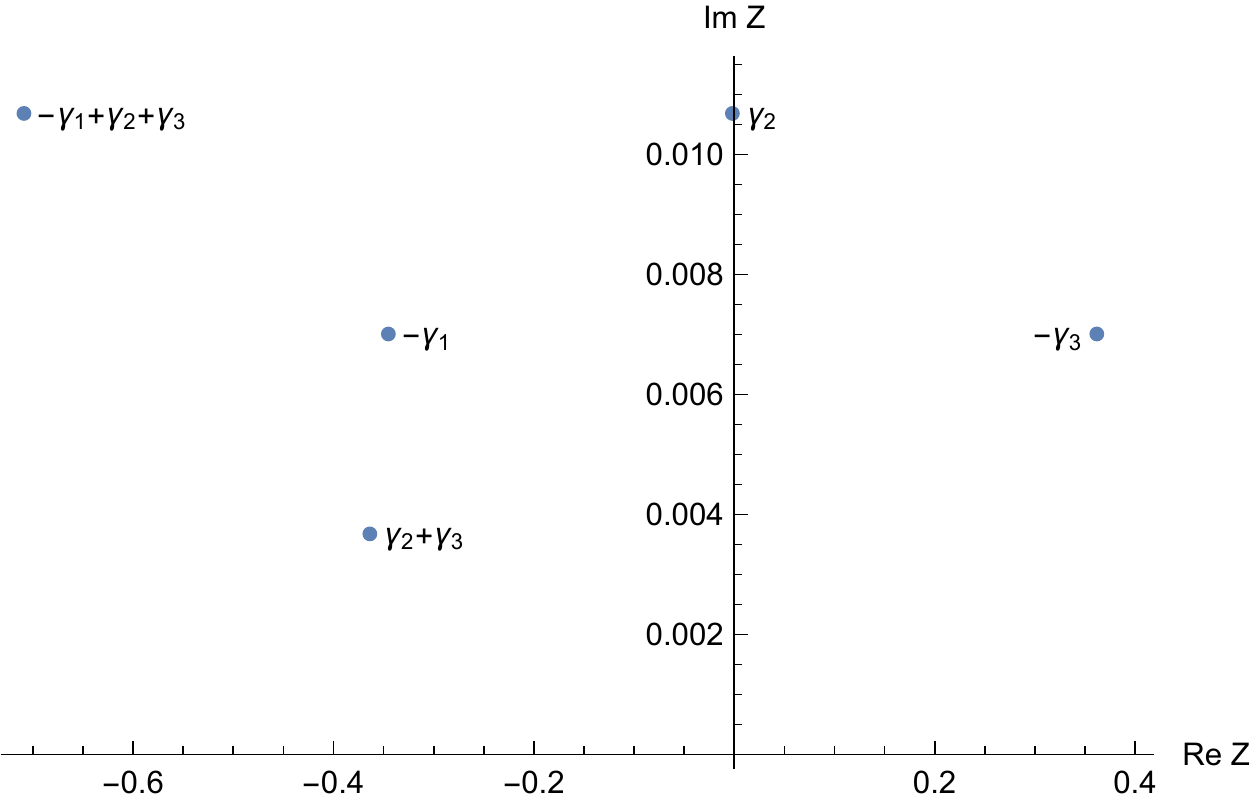}\\\\
\includegraphics[scale = 0.6]{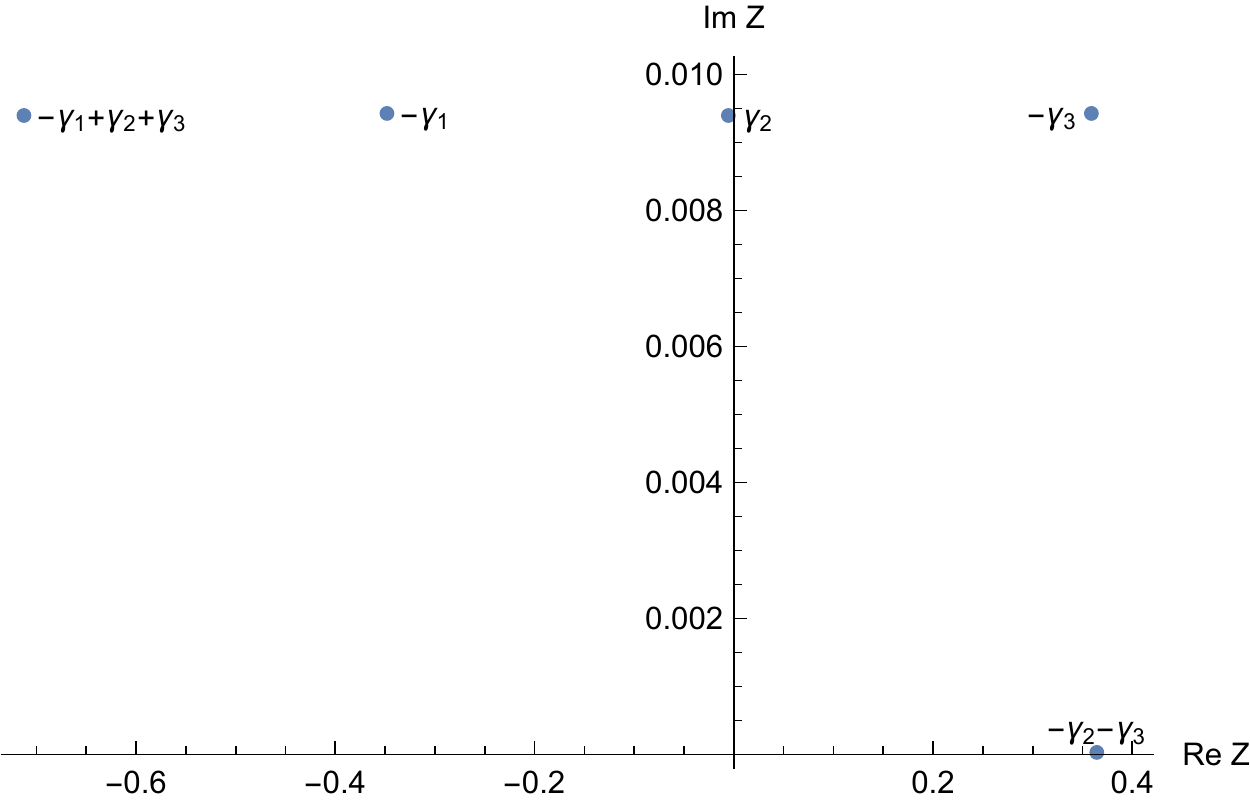}& &
\includegraphics[scale = 0.6]{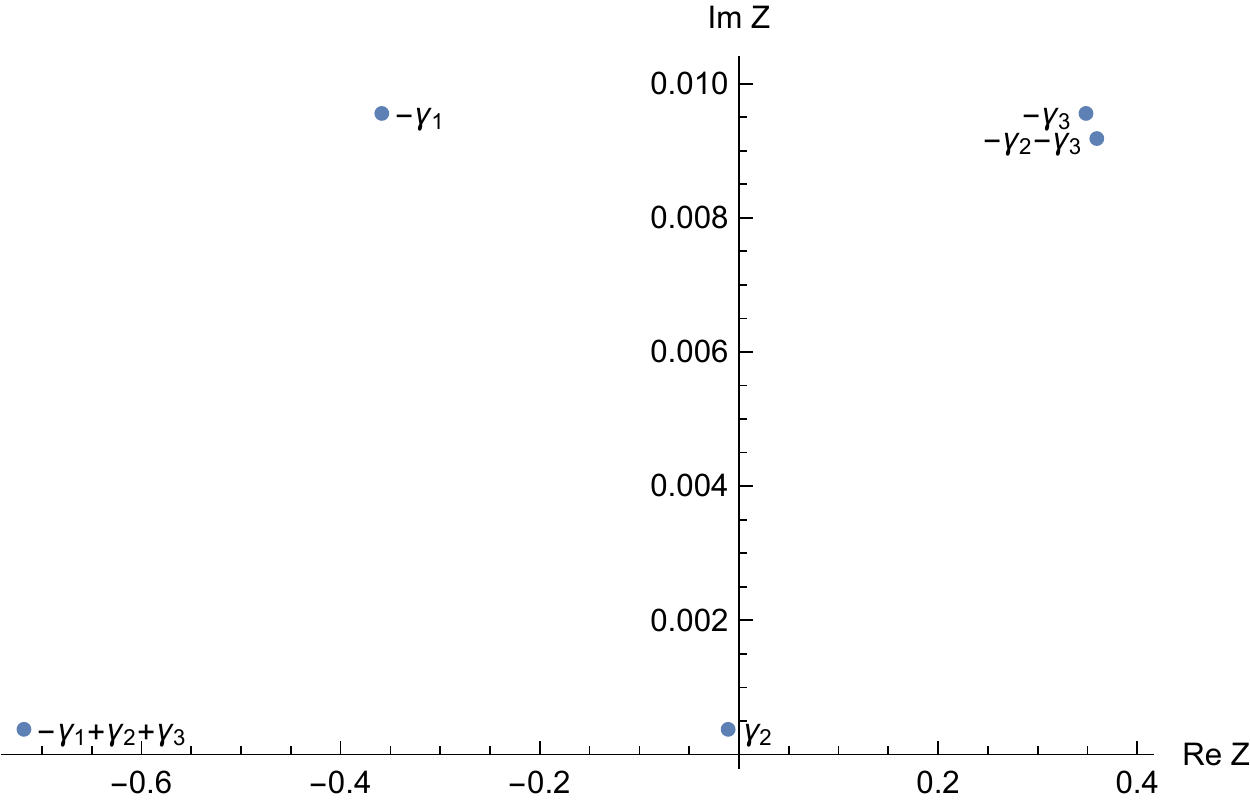}\\\\
\end{tabular}
\caption{\textsc{up left:} Stable states with charges $\gamma_1,\gamma_2,\gamma_3, \gamma_2+\gamma_3$
for $\theta=5.3$.
\textsc{up right:} double mutation at $\theta=2\pi$.
\textsc{center left:} new quiver basis valid for $0<\theta<1.05$.
\textsc{center right:} the wall crossing at which the state of charge $\gamma_2 + \gamma_3-\gamma_1$ enters in the spectrum occurs at $\theta = 0.525$, here we plot the stable states in the Z-plane at $\theta=0.7$.
\textsc{down left:} Right after the $\theta \approx 1.05$ mutation at $\gamma_2 + \gamma_3$.
\textsc{down right:} $\theta \approx 2.06$ right before the double mutation at $-\gamma_1 + \gamma_2 + \gamma_3$ and $\gamma_2$ and the wall-crossing leading to the disappearence of $\gamma_3$ from the spectrum and the appearance of $\gamma_1 + \gamma_2$.}\label{WXlargeMpatty1}
\end{figure}

\begin{figure}
\begin{tabular}{ccc}
\includegraphics[scale = 0.6]{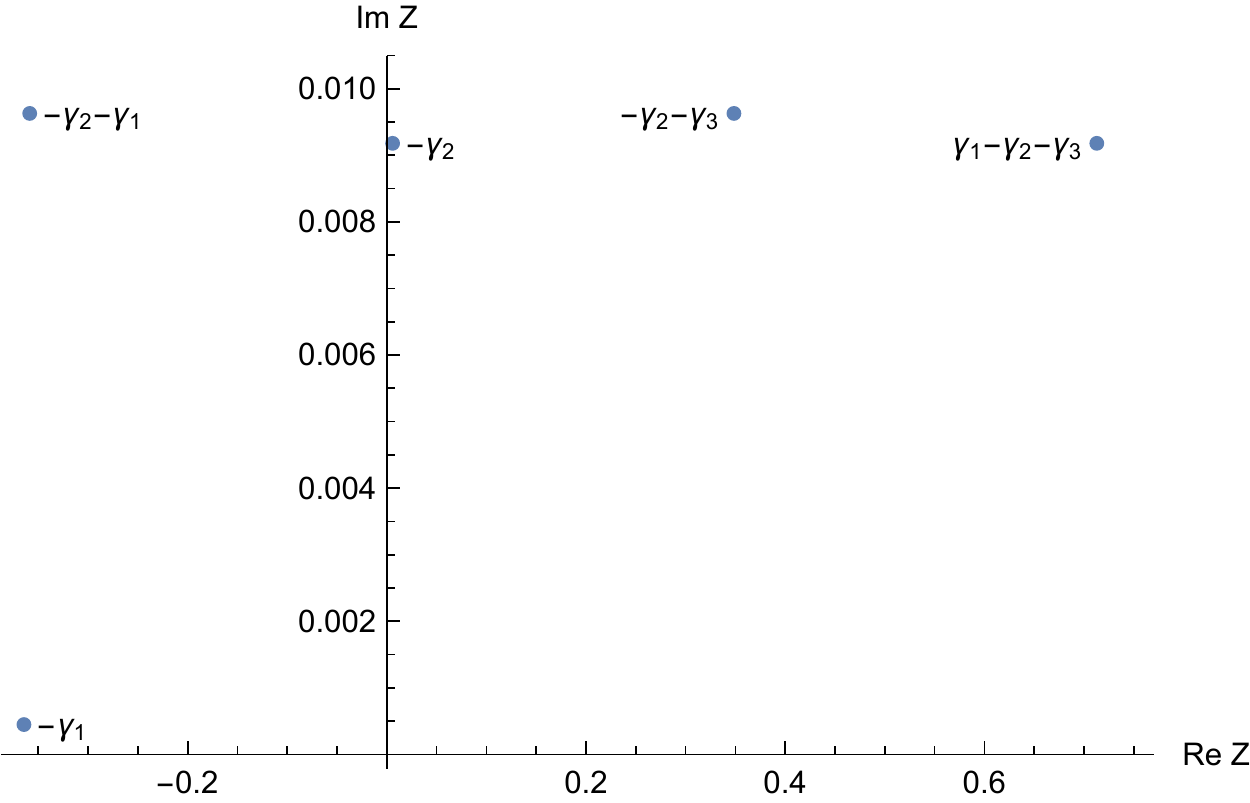}& &
\includegraphics[scale = 0.6]{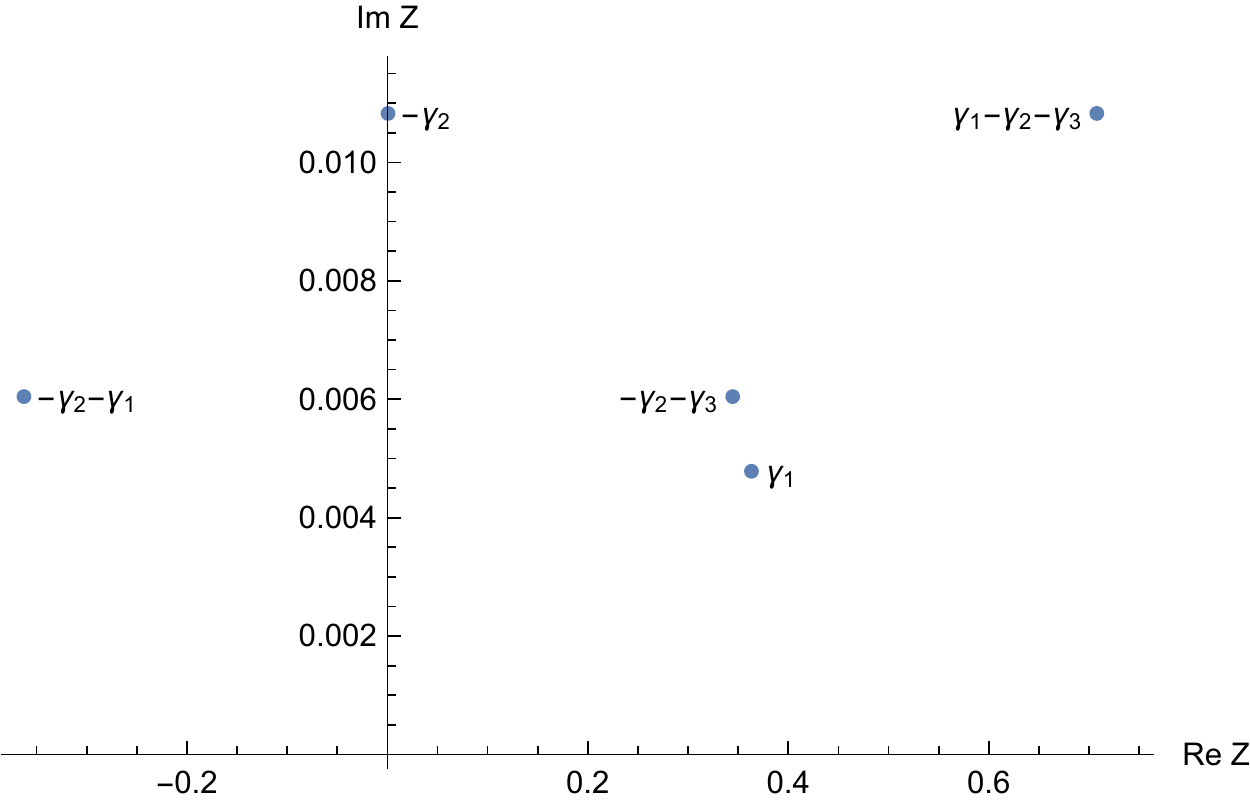}\\\\
\includegraphics[scale = 0.6]{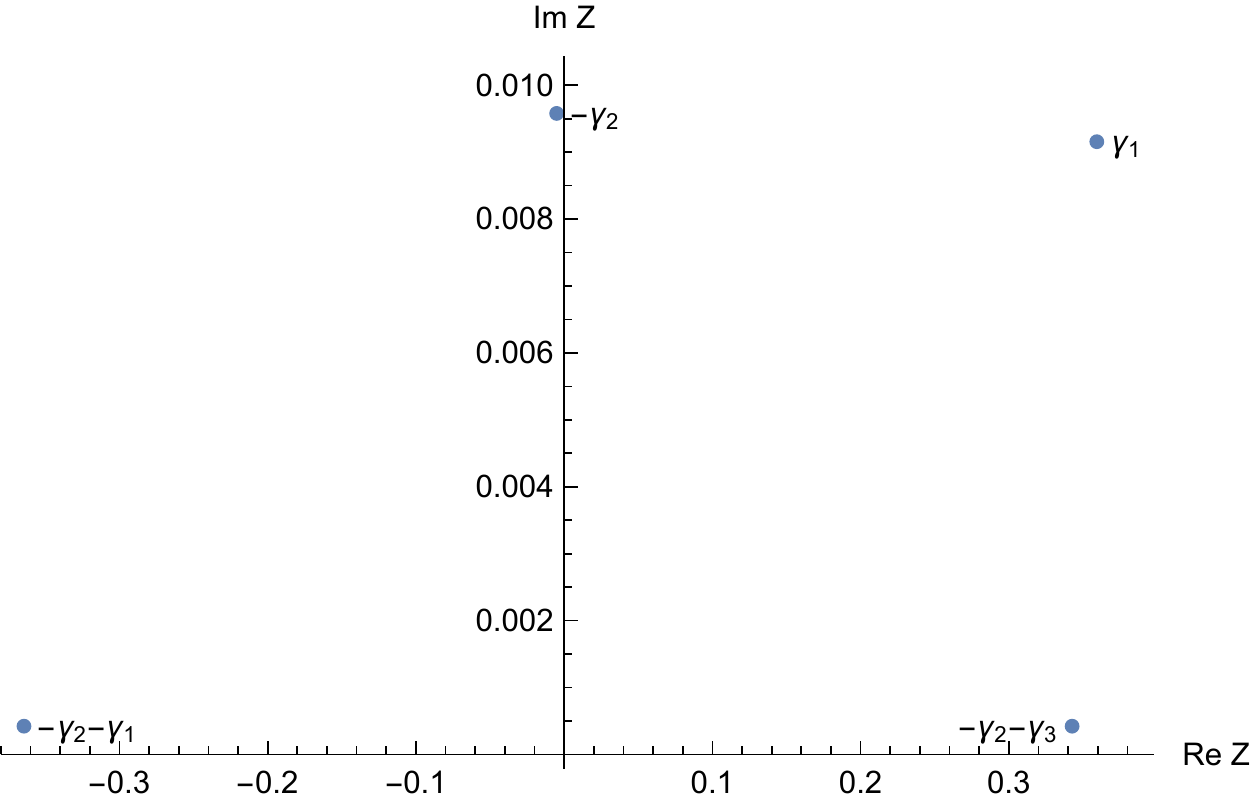}& &
\includegraphics[scale = 0.6]{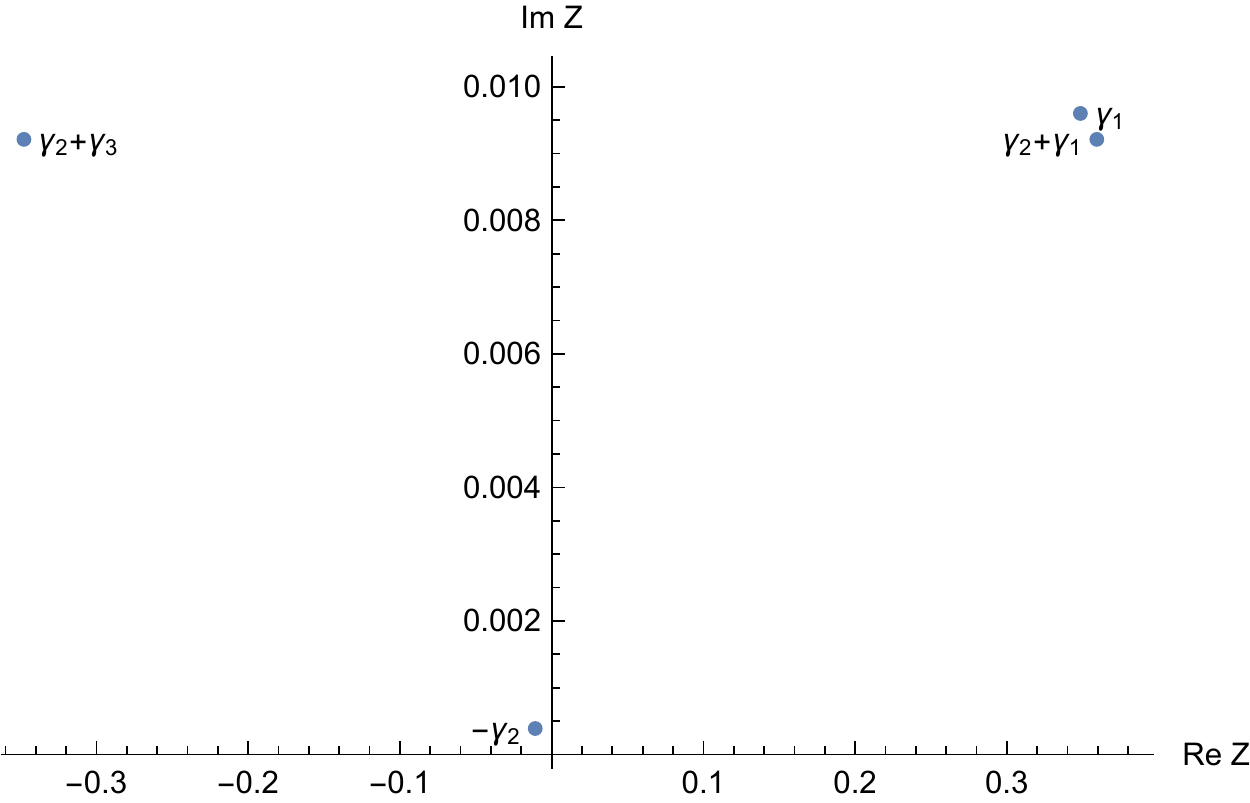}\\\\
\end{tabular}
\caption{\textsc{up left:} $\theta \approx 3.1$ right before the mutation at $\pi$.
\textsc{up right:} $\theta \approx 3.6$ right before the wall-crossing leading to the decay of the state  $\gamma_1 - \gamma_2 - \gamma_3$.
\textsc{down left:} Right before the mutation at $\theta \approx 4.15$.
\textsc{down right:} Right before the wall-crossings destabilizing $\gamma_1 + \gamma_2$ and stabilizing $\gamma_3$ while $\gamma_2$ mutates at $\theta \approx 5.2$}\label{WXlargeMpatty2}
\end{figure}

For $5 \pi/3 < \theta <2 \pi$, the quiver basis we start with gives a BPS quiver
$$\gamma_1 \longrightarrow \gamma_2 \longleftarrow \gamma_3.$$
The corresponding central charges and stable states are depicted in figure \ref{WXlargeMpatty1},
we have BPS spectrum
\begin{equation}\gamma_1,\gamma_2,\gamma_3,\gamma_2 + \gamma_3, \text{ and CPT conjugates}\label{largeMspeK1}\end{equation}
At $\theta = 2 \pi$ a double mutation occurs (see figure \ref{WXlargeMpatty1}):
$Z(\gamma_1)$ and $Z(\gamma_3)$ exits the upper $Z$-plane simultaneously. The mutated quiver is
\begin{equation}-\gamma_1\longleftarrow\gamma_2 +\gamma_3 \longrightarrow  -\gamma_3.\label{largemassQ2}\end{equation}
The BPS spectrum remains the one in \eqref{largeMspeK1}, the stable particles have charges
$-\gamma_1, \gamma_2 + \gamma_3, -\gamma_3,\gamma_2$.
Now $\gamma_2$ appears as a stable bound state with dimension vector $(0,1,1)$ for the
$A_3$ quiver in line \ref{largemassQ2}, adding CPT conjugates one gets the same charges
as in line \eqref{largeMspeK1}. At $\theta \sim 0.525$ a wall-crossing phase transition occurs and the BPS state with charge $\gamma_2 +\gamma_3 -
\gamma_1$ stabilizes: the BPS spectrum becomes
\begin{equation}-\gamma_1,\gamma_2 + \gamma_3,-\gamma_3,\gamma_2, \gamma_2 +\gamma_3 -\gamma_1\text{ and CPT conjugates}.
\label{largeMspeK2}\end{equation}
At $\theta\approx1.05$ another mutation occurs (see figure \ref{WXlargeMpatty1}) the charge $\gamma_2 + \gamma_3$ exits the upper half Z-plane and the BPS quiver becomes
\begin{equation} \gamma_2 + \gamma_3 -\gamma_1\longrightarrow - (\gamma_2 + \gamma_3) \longleftarrow  -
\gamma_2.\label{largemassQ3}\end{equation}
The spectrum is still as in line \eqref{largeMspeK2}, but now $\gamma_3$
is a bound state corresponding to the dimension vector $(0,1,1)$. At $\theta\approx2.06$ both the charge $\gamma_2 + \gamma_3 -\gamma_1$ and the
charge $\gamma_2$ exit the upper half Z-plane. Moreover, the state with charge $\gamma_3$ destabilizes while and the state with charge $-
(\gamma_1+\gamma_2)$ stabilizes (see figure \ref{WXlargeMpatty1}). The mutated BPS quiver is
\begin{equation} - \gamma_2 - \gamma_3 +\gamma_1  \longleftarrow - \gamma_1 \longrightarrow  -\gamma_2,\label{largemassQ4}\end{equation}
The new BPS spectrum is
\begin{equation} - \gamma_2 - \gamma_3 + \gamma_1, -\gamma_1,-\gamma_2, -\gamma_2 - \gamma_3,-\gamma_1-\gamma_2\text{ and CPT
conjugates}.\label{largeMspeK3}\end{equation}
At $\theta = \pi$  (see figure \ref{WXlargeMpatty2}) the charge $\gamma_1$ mutates and one has the quiver
\begin{equation} - \gamma_2 - \gamma_3  \longrightarrow  \gamma_1  \longleftarrow  -\gamma_2 - \gamma_1,\label{largemassQ5}\end{equation}
At $\theta \approx 3.65$ there is a wall-crossing and the spectrum becomes:
\begin{equation}\gamma_1,-\gamma_2, -\gamma_2 - \gamma_3,-\gamma_1-\gamma_2\text{ and CPT
conjugates}.\label{largeMspeK4}\end{equation}
At $\theta \approx 4.15$ we have another double mutation (see figure \ref{WXlargeMpatty2})
\begin{equation} \gamma_2 + \gamma_3  \longleftarrow  -\gamma_2 \longrightarrow  \gamma_2+ \gamma_1,\label{largemassQ5}\end{equation}
followed at $\theta \approx 5\pi/3$ by a mutation at $\gamma_2$ and simultaneously two wall-crossings leading to the destabilization of the state with charge $\gamma_2+ \gamma_1$, and the stabilization of the state with charge $\gamma_3$, bringing us back to the original spectrum at $\theta > 5 \pi /3$ (see figure \ref{WXlargeMpatty2}).

\medskip

\begin{figure}
\begin{tabular}{ccc}
\includegraphics[scale = 0.6]{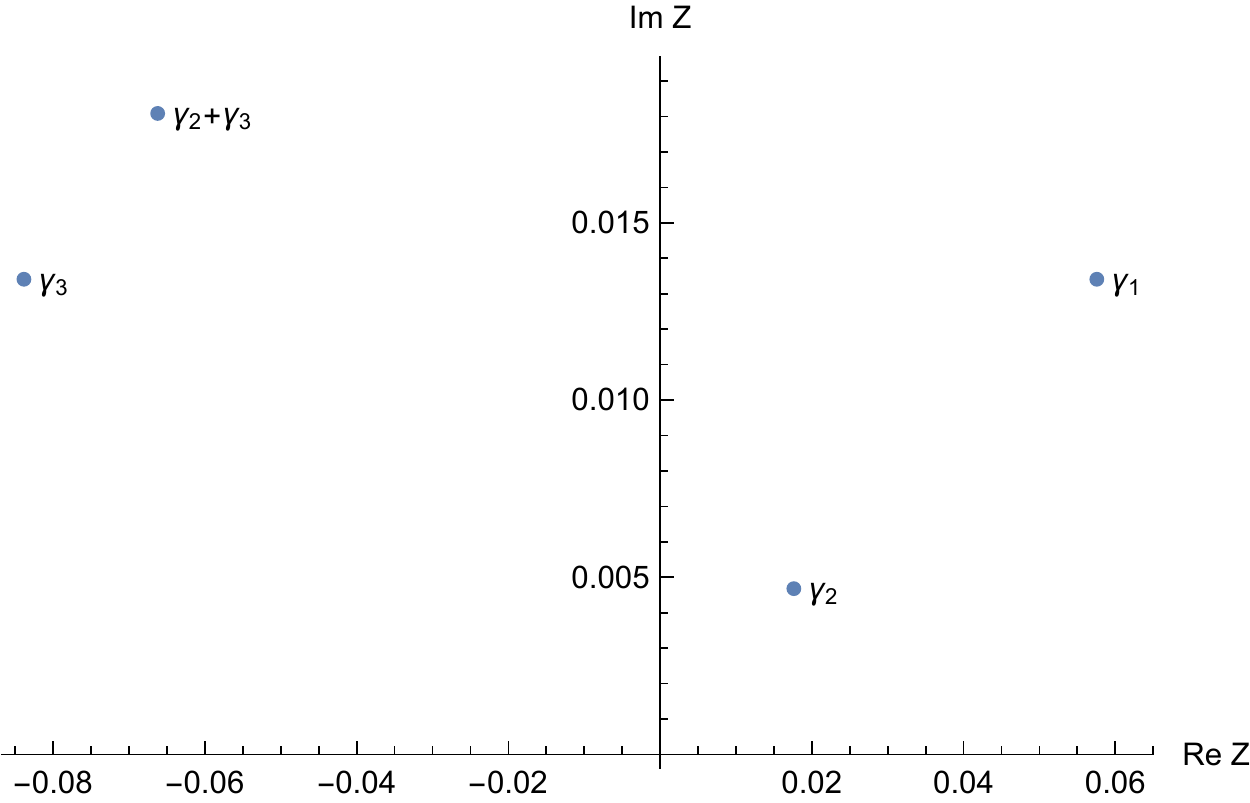}& &
\includegraphics[scale = 0.6]{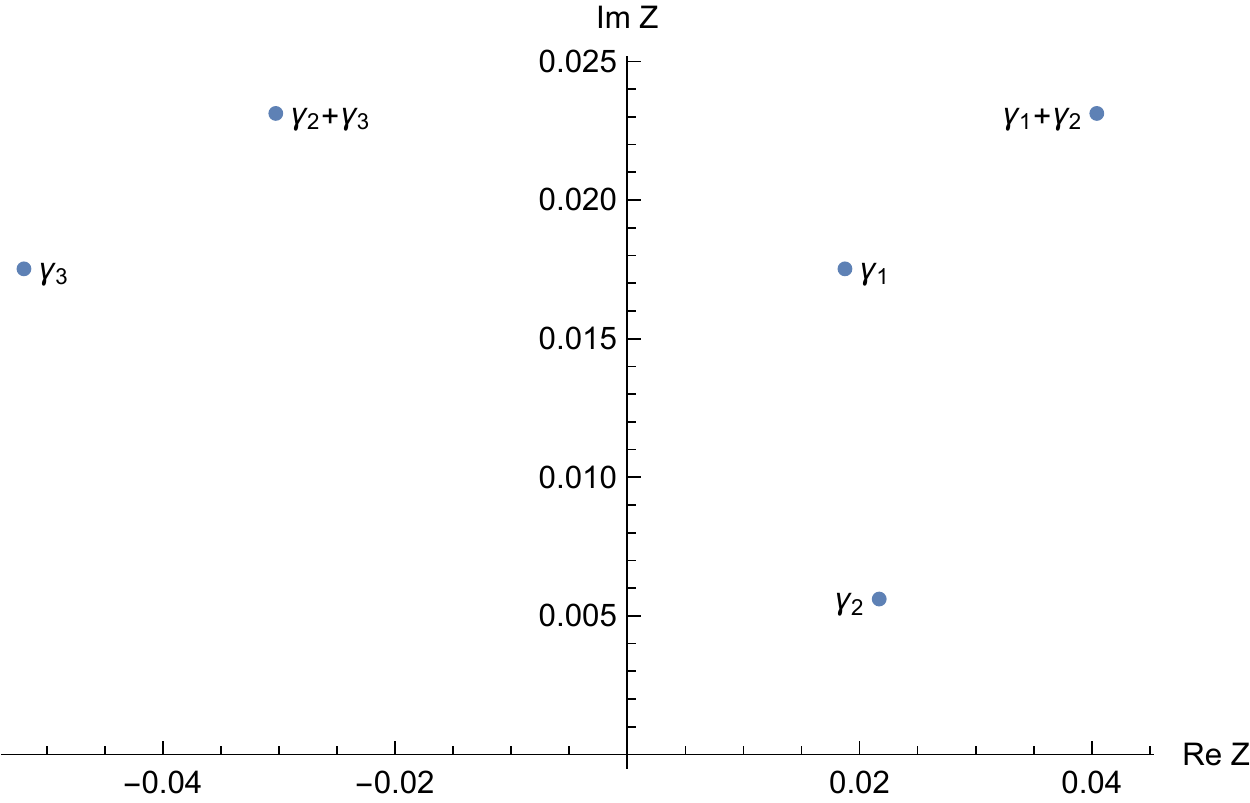}\\\\
\includegraphics[scale = 0.6]{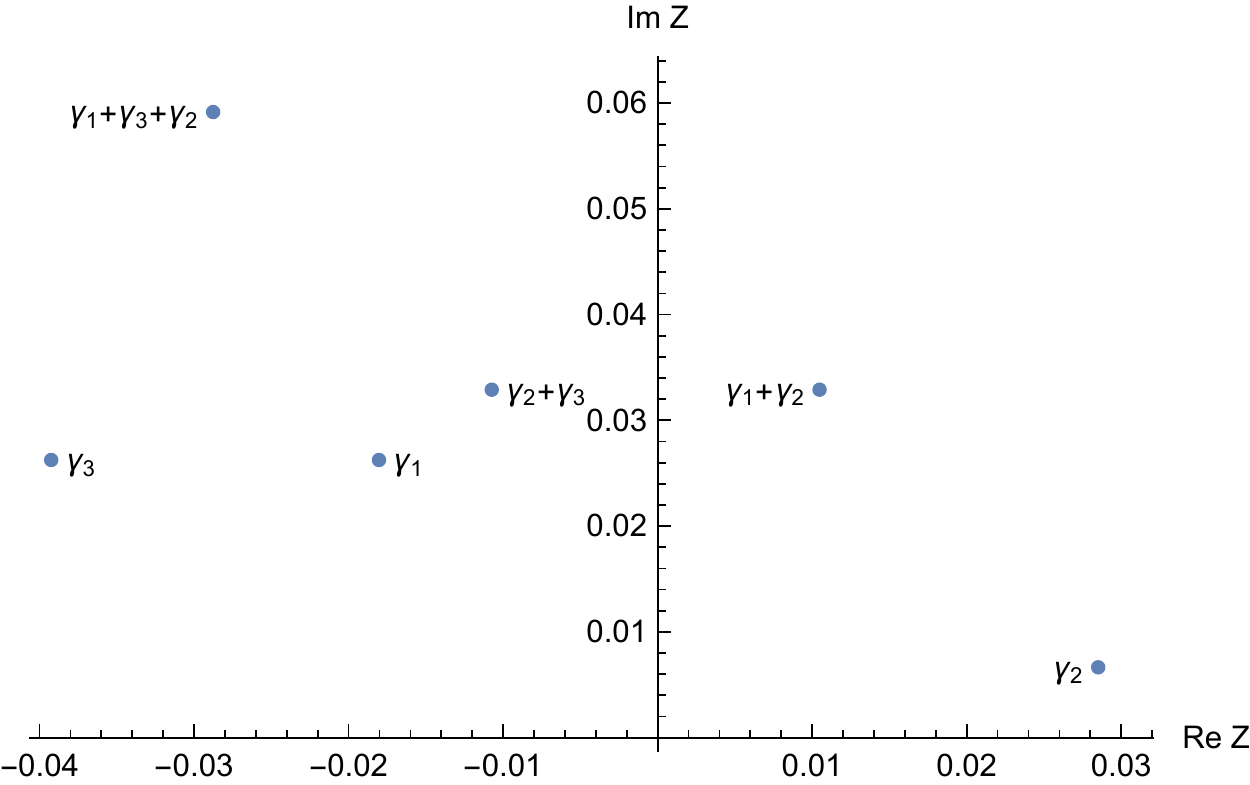}& &
\includegraphics[scale = 0.6]{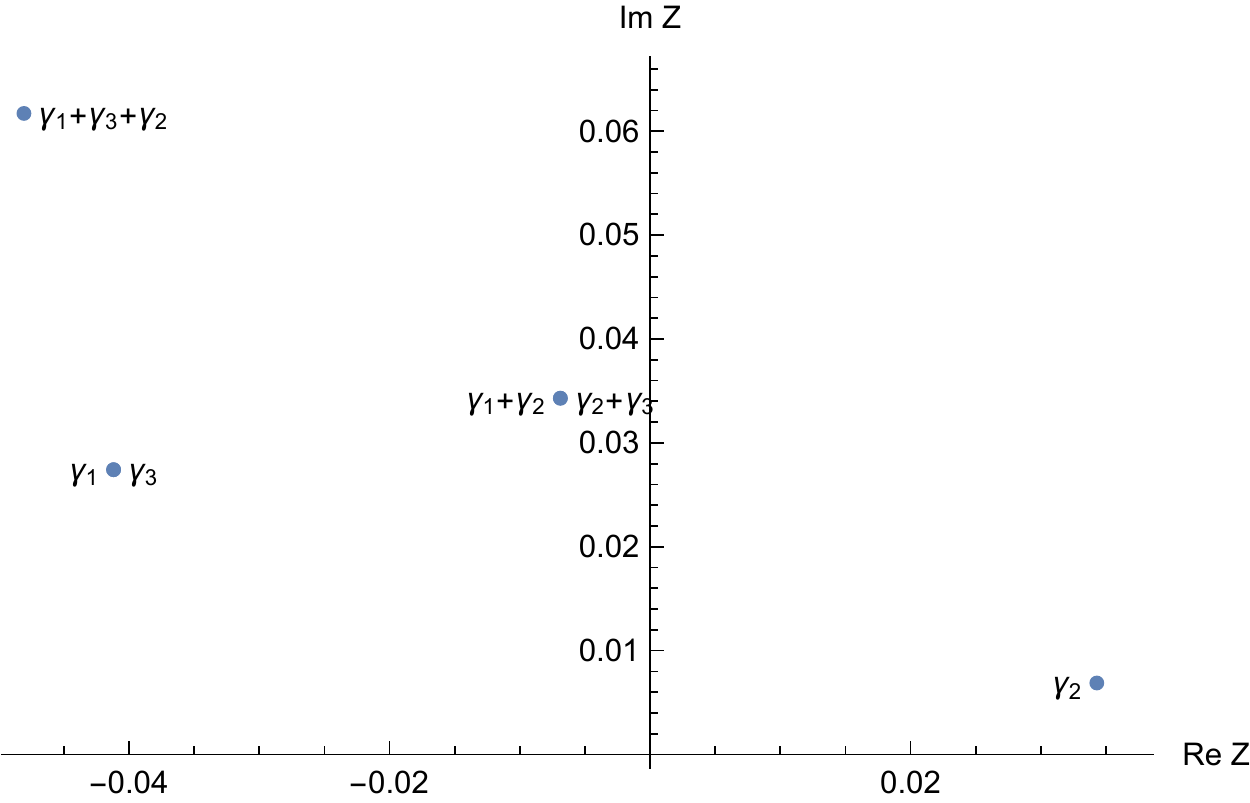}\\\\
\end{tabular}
\caption{\textsc{up left:} $m \approx 0.2$ same spectrum as at large mass for $\theta = 5.5$.
\textsc{up right:} $m \approx 0.1$ the state $\gamma_1 + \gamma_2$ enters the spectrum.
\textsc{down left:}  $m \approx 0.03$ right after the mutation leading to the stabilization of the dark dyon.
\textsc{down right:} $m \approx 0$ the flavor symmetry gets restored.}\label{WXsmallMpatty1}
\end{figure}

The BPS spectrum in figure \ref{smalltolargemass} was obtained for $u=0.1e^{i5.5}$ tuning $m$ from $0$ to $1$. The whole line is covered by the quiver
\begin{equation}
\gamma_1 \longrightarrow \gamma_2 \longleftarrow \gamma_3.
\end{equation}
The relevant pattern of wall-crossings in the $Z$-plane is illustrated in figure \ref{WXsmallMpatty1}: for $0.2 < m < 1$ the spectrum is constant. In the region $0.03 < m < 0.2$ a series of wall crossing occurs leading to a maximal chamber with stable states
\begin{equation}\gamma_1,\gamma_2,\gamma_3,\gamma_1 + \gamma_2 , \gamma_2+\gamma_3,\gamma_1+\gamma_2+\gamma_3\end{equation}

\medskip

It should be possible to reproduce our results using the spectral networks as in \cite{Maruyoshi:2013fwa}.

\newpage

\bibliographystyle{utphys}
\bibliography{MagMix}

\end{document}